\documentclass[namedreferences]{kluwer}    %
\usepackage{aastexjournals}
\usepackage{graphicx}

\def\deg{\hbox{$^\circ$}}

\def\mn{\hbox{$\mu$m}}
\def\iras{{\em IRAS}}
\def\iso{{\em ISO}}
\def\sws{{\em ISOSWS}}
\def\lws{{\em ISOLWS}}
\def\pht{{\em ISOPHOT}}
\def\cam{{\em ISOCAM}}
\def\hl{HyLIG}
\def\ul{ULIG}
\def\lg{LIG}
\def\hls{HyLIGs}
\def\uls{ULIGs}
\def\ls{LIGs}
\def\ir{infrared}
\def\mir{mid-infrared}
\def\fir{far-infrared}
\def\nir{near-infrared}

\def\fsl{fine structure line}
\def\hrl{H-recombination line}
\def\n{NGC\,}
\def\sf{star formation}
\def\wr{Wolf-Rayet}

\def\hii{H\,II}

\def\lsun{$L_{\odot}$}

\newdisplay{guess}{Conjecture}

\begin{document}                                                                                   

\begin{article}

\begin{opening}         

  \title{Obscured Activity: AGN, Quasars, Starbursts and ULIGs
    observed by the Infrared Space Observatory \thanks{Based
          on observations with ISO, an ESA project
          with instruments funded by ESA Member States (especially
          the PI countries: France, Germany, the Netherlands
          and the United Kingdom) and with the participation
          of ISAS and NASA.}}

  \author{Aprajita \surname{Verma}\email{verma@mpe.mpg.de}}
  \institute{Max-Planck Institut f\"ur extraterrestrische Physik,
    Postfach 1312, D-85741 Garching, Germany.} 

  \author{Vassilis
  \surname{Charmandaris}\email{vassilis@astro.cornell.edu}}
  \institute{Department of Astronomy, Cornell University, 106 Space
  Sciences Building, Ithaca, NY 14853.  Chercheur Associ\'e,
  Observatoire de Paris, F-75014, Paris, France.  Now at the University of
  Crete, Department of Physics, GR-71003, Heraklion, Greece.}

  \author{Ulrich \surname{Klaas}\email{klaas@mpia-hd.mpg.de}}
  \institute{Max-Planck-Institut f\"ur Astronomie, K\"onigstuhl 17, D-69117
    Heidelberg, Germany.} 

  \author{Dieter \surname{Lutz}\email{lutz@mpe.mpg.de}}
  \institute{Max-Planck Institut f\"ur extraterrestrische Physik,
    Postfach 1312, D-85741 Garching, Germany.} 

  \author{Martin \surname{Haas}\email{haas@astro.ruhr-uni-bochum.de}}
  \institute{Astronomisches Institut, Ruhr-Universit\"at,
    Universit\"atsstr. 150, D-44780 Bochum, Germany.}

  \runningauthor{Verma et al.}
  \runningtitle{Obscured ``Activity'': AGN, Quasars, Starbursts and
  ULIGs}

%\date{16th May 2004}

\begin{abstract}
Some of the most 'active' galaxies in the Universe are obscured by large
quantities of dust and emit a substantial fraction of their bolometric
luminosity in the \ir. Observations of these infrared luminous galaxies
with the Infrared Space Observatory (\iso) have provided a relatively
unabsorbed view to the sources fuelling this active emission. The 
improved sensitivity, spatial resolution and spectroscopic capability of 
\iso\ over its predecessor Infrared Astronomical Satellite (\iras), 
has enabled significant advances in the understanding of the \ir\ properties 
of active galaxies.  \iso\ surveyed a wide range of active galaxies 
which, in the context of this review, includes those powered by intense 
bursts of star-formation as well as those containing a dominant active 
galactic nucleus (AGN). Mid infrared imaging resolved for the first time 
the dust enshrouded nuclei in many nearby galaxies, while a new era in 
\ir\ spectroscopy was opened by probing a wealth of atomic, ionic and 
molecular lines as well as broad band features in the mid and far 
infrared. This was particularly  useful since it resulted in the 
understanding of the power production, excitation and fuelling mechanisms 
in the nuclei of active galaxies including the intriguing but so far 
elusive ultraluminous infrared galaxies. Detailed studies of various 
classes of AGN and quasars greatly improved our understanding of the 
unification scenario. Far-\ir\ imaging and photometry also revealed the 
presence of a new very cold dust component in galaxies and furthered our 
knowledge of the \fir\ properties of faint starbursts, \uls\ and 
quasars. We summarise almost nine years of key results based upon \iso\ 
data spanning the full range of luminosity and type of active galaxies.

\end{abstract}


\end{opening}
\noindent{\bf Received: } 4 November 2004  {\bf Accepted: } 15 November 2004 

\section{Introduction}
\label{intro}

``Activity'' is a common feature of galaxies that emit a large fraction of their energy in the
infrared.  Existing primarily in a state of quiescence, 
during certain periods of their evolution galaxies experience short-lived phases 
of extreme star formation and/or black hole
activity.
From early ground-based measurements (e.g.
\opencite{1968AJ.....73..868Low};
\citeauthor{1970ApJ...159L.165Kleinmann}
\citeyear{1970ApJ...159L.165Kleinmann,1970ApJ...161L.203Kleinmann}) to
surveys performed by the Infrared Astronomical Satellite (\iras,
\opencite{1988IRAS..C......0Beichman}), many previously known active
galaxies [e.g. optical- or radio-selected quasars, active galactic
nuclei (AGN), and starbursts] were found to be as or more powerful in the
infrared than in the visible. In addition, a new population of optically faint, infrared
luminous galaxies (luminous infrared galaxies (\ls, 
with $L_{IR}>10^{11}L_{\odot}$) were discovered that emit most of their
bolometric luminosity in the infrared \cite{1984ApJ...278L..63Houck}. AGN and massive bursts of star
formation are the only known mechanisms that are capable of generating
such luminosities. However, without spectroscopic information of the
obscured components, classifying the underlying power source has proved to be 
difficult.
Furthermore, quantitatively assessing the contribution to the total power from the
two possible mechanisms is non-trivial.  It is known that
dust is formed during the late stages of stellar evolution and it is
destroyed by intense radiation fields and shocks. Consequently, a
detailed understanding of the physical mechanisms responsible for the
observed \ir\ emission is an essential first step into revealing the
nature of the underlying exciting source. As a result, since the most
'active' regions in our Universe appear to be enshrouded in large
quantities of gas and dust, addressing the aforementioned points would have
serious consequences in estimating star formation and black hole
activity in the Universe.

In a cosmological context, both the reported increase
in the star formation density of the Universe for $0.1\lsim z \lsim
2$ (e.g. \opencite{1996MNRAS.283.1388Madau};
\opencite{2000AJ....119.2092Barger}) and the high frequency of
starbursts hosted in galaxies with disturbed morphologies or in
interacting/merging systems, imply that starbursts
play an important role in galaxy formation and
evolution. In addition, black hole activity could be buried in any
luminous \ir\ system. 
Ultraluminous infrared galaxies (\uls, $L_{IR}>10^{12}L_{\odot}$) have
been proposed to be an evolutionary stage in the life of a quasar and
the local analogues of the $z>2$ (sub-)mm population.
Quantifying the amount of
obscured activity at earlier times and its contribution to the
infrared and X-ray backgrounds remain open issues requiring a
sound understanding of the properties of local active galaxies.

Following on from the coarse, photometric
legacy of \iras, \iso\ provided the first means to investigate the
physical conditions of local active galaxies in significant detail at
wavelengths where the obscured, and often most active, regions could
be probed. Equipped with imaging, photometric and spectroscopic
capabilities, \iso\ surveyed a wide range of emission properties from
broad-band spectral energy distributions (SEDs), through imaging, to
detailed spectral analysis.  We discuss the infrared emission
properties of obscured active galaxies that \iso\ surveyed during its
lifetime. In the context of this chapter, 'active' refers to both star
formation and black hole activity and encompasses the range of
non-normal galaxies: interacting/merging galaxies;
starbursts; radio galaxies; AGN; quasars; low ionisation nebular
emission regions (LINERs); and the full suite of \ls\ - \uls\
and hyperluminous infrared galaxies (\hls\ $>10^{13}$\lsun).  

\section{Activity manifest in the infrared}
\label{iractivity}

High energy UV (and visible) photons emanating from
active regions 
can heat or excite environmental dust, located around active sites or
distributed throughout the interstellar medium (ISM), resulting
in re-radiation in the \ir.
Grains ranging from the very small (radius a$\lsim$10nm)
to the large (a$\sim$30\mn) contribute to emission over the
2.5-200\mn\ range. The type, size and distribution of dust grains
together with the incident interstellar radiation field (ISRF) shape
the observed spectral features and overall
form of the SED which arise from both thermal and non-thermal processes.
At long \ir\ wavelengths ($\lambda\gsim$25\mn), emission predominantly
arises from grains that re-radiate in thermodynamic equilibrium.  In
young radio galaxies, thermal bremsstrahlung may also provide a
contribution to the \ir\ continuum. Emission related to transient
rather than steady-state heating of dust grain complexes can dominate
the SEDs of some active galaxies giving rise to continua and features
over 3-18\mn.
Non-thermal components include the featureless \fir\ tail (increases
with increasing wavelength) of the synchrotron radiation spectrum in
strong radio galaxies.

\subsection{Infrared continua: thermal and non-thermal}

 The \mir\ spectra of galaxies display two continua:
a flat continuum for $\lambda
\lsim$5\mn\ and, in some galaxies, an unrelated continuum for $\lambda
\gsim$11\mn\ \cite{2000ApJ...532L..21Helou}. 
These continua are not produced by grains heated to thermal
equilibrium by the radiation field
but arise from the stochastic heating of small dust grains
(amorphous silicates and graphites, a$\sim$0.01-0.1\mn) which are
transiently heated by the absorption of a single photon
\cite{1984ApJ...277..623Sellgren,1987ARA&A..25..521Beichman,1998A&A...339..194Boulanger,2001ApJ...551..807Draine,2003ARA&A..41..241Draine}.
A single photon with energy of only a few electron volts ($\lambda \lsim$0.4\mn), is
sufficient for the onset of infrared emission
\cite{1998ApJ...493L.109Uchida}.

The steeply rising (near-thermal) continuum longward of 11\mn, which is strong in
some active galaxies (but weak in quiescent galaxies), is thought to originate from the excess
transient heating of very small fluctuating dust grains (VSG,
a$\le10$nm). This component appears to be a characteristic of intensely
star forming regions 
\cite{1990A&A...237..215Desert,1996A&A...315L.337Verstraete,1996A&A...315L.305Cesarsky,1996A&A...315L.313Laureijs}. 
Under this transient heating regime, the mid-infrared continuum is
directly proportional to the underlying radiation field over several
orders of magnitude
and varies with the conditions of the \hii\ regions from which it
originates \cite{1998sfis.conf...15Boulanger,2000A&A...359..887Laurent,2004A&A...419..501Foerster}.  For example, in the close vicinity of the dense starburst
knots in the Antennae, this continuum is shifted to shorter \mir\
wavelengths implying higher temperatures \cite{1996A&A...315L..93Vigroux}.

The 3\mn$<\lambda<$5\mn\ continuum is likely to arise from a featureless
fluctuating component \cite{2000ApJ...532L..21Helou}. Although often weak in active galaxies, the direct stellar continuum can also contribute ($\lambda<$5\mn) and should be considered in careful spectral decomposition analysis. 
The strength of this component increases with decreasing wavelength and thus the transition between direct stellar to interstellar
dust emission occurs within the near- to low-\mir\
\cite{1997A&A...324L..13Boselli,1998A&A...335...53Boselli,2001AJ....121.1369Alonso}. The \mir\ also includes the transition from the transient to
the steady-state heating regime. The wavelength at which this
transition occurs is determined by the incident
radiation field above which grains of a certain size no longer suffer
large temperature changes.
Dust grains 
are heated by the general ISRF and attain a
'steady-state' where emission and absorption reach an equilibrium. The
grain temperature at which this equilibrium occurs characterises the
emission giving rise to a thermal black-body continuum.
Qualitatively, the mid- to far-\ir\ continua of active galaxies are often
approximated by a series of superposed black-bodies, each originating
from a body of dust heated to a characteristic temperature
\cite{1998ApJ...503L.109Haas,1998ApJ...494..211Ivison,2001A&A...379..823Klaas,2003AJ....125.2361Bendo,2005MNRAS.357..361Stevens}. This
approximation provides only an indication of typical temperatures and
is not an accurate description of the complex heating of a variety of
dust grains. More accurate descriptions are achieved by considering
a multi-grain dust model and complex radiation fields using, for example,
semi-empirical models (e.g. \opencite{2001ApJ...549..215Dale}) or radiative
transfer theory
(e.g. \opencite{1996A&A...315L.121Acosta}; \opencite{1998ApJ...509..103Silva}; \opencite{1999MNRAS.310...78Alexander}; \opencite{1999A&A...351..140Siebenmorgen}; \opencite{2000MNRAS.313..734Efstathiou}; \opencite{2002MNRAS.335..574Verma}; \opencite{2003MNRAS.343..585Farrah}; \opencite{2003ApJ...599L..13Freudling}; \opencite{2004ApJ...613..247Gonzalez}; \opencite{2004A&A...421..129Siebenmorgen}). For
sources that are spatially unresolved by the \iso\ instruments, SED decomposition (using one of the methods above) is the only means to investigate the underlying physical processes.

Some AGN-hosting galaxies display a relatively
broad and flat (in $\nu F_{\nu}$) hot continuum that dominates their \mir\ SEDs. This emerges from grains in the putative torus that are heated up to the grain sublimation temperature (T$_{sub}\sim$1500K) by the strong radiation field of the accretion disk of the central AGN. 
The 'warm' \iras\ galaxy
criterion S$_{60}$/S$_{25}>0.2$) selects galaxies displaying this
component which are mostly identified with AGN \cite{1985Natur.314..240Degrijp}.

The extended wavelength range of \iso\ over \iras\ enabled (a) the precise determination of the 
turnover in the \fir\ SEDs of active galaxies and
(b) the identification of two dust components that comprise the \fir\ emission that are 
both associated with large grains in thermal equilibrium.
The first (or 'cold') component represents dust surrounding star
forming regions and thus its strength is directly related to the star
formation rate (however see Sect. \ref{sfr}). While the SEDs of normal
quiescent galaxies peak at $\sim$150\mn, the enhanced heating in
actively star forming regions produces a strong \fir\ peak at
$\sim$60-100\mn\ that dominates the \ir\ SED and corresponds to dust
heated to 30-60K (e.g. \opencite{2000ApJ...533..682Calzetti};
\opencite{2001ApJ...549..215Dale};
\opencite{2002ApJ...572..105Spinoglio}). This component also enhances
the \mir\ continuum (e.g. see Figure 1 in
\opencite{1996ARA&A..34..749Sanders}).  In AGN, the fractional
contribution of this 'cold' \fir\ component can be weaker in
comparison to the warmer \mir\ component ($\lambda \sim 30$\mn) than
in starbursts, reflective of the intense heating power of the AGN.
The second ('very cold' or 'cirrus') component is dust that is spatially more
extended and
is associated with the
HI disk or the molecular halo \cite{2000ApJ...543..153Trewhella,2001A&A...377...73Radovich,2004A&A...415...95Stickel}. Located far 
and, in some galaxies, shielded \cite{2000ApJ...533..682Calzetti} from the intense emission from the central engine, the extended component is heated
by the ambient
interstellar radiation field to temperatures of 10-20K,
\cite{1998ApJ...500..554Odenwald,1998A&A...335..807Alton,1999MNRAS.304..495Davies,1998A&A...338L..33Haas,1998A&A...337L...1Haas,2000ApJ...543..153Trewhella}. These temperatures are consistent with the theoretically predicted heating/cooling of dust
grains in thermal equilibrium by a diffuse radiation field.
Both
profile measurements for resolved sources and \fir-sub-mm SED analysis for more distant galaxies
reveal that a very
cold component is required to explain all the emission
beyond 100\mn\
\cite{1996A&A...315L.129Rodriguez,1999A&A...348..705Radovich,1999A&A...351..495Siebenmorgen,1999AJ....118.1542Domingue,2000ApJ...533..682Calzetti,2000A&A...356L..83Haas,2001A&A...379..823Klaas,2001ApJ...557...39Perez,2003A&A...402...87Haas}. This
component, that \iras\ was insensitive to, may be a significant
contributor to the \fir\ luminosity in some active galaxies
(e.g. $\gsim 60\%$ in local starbursts
\opencite{2000ApJ...533..682Calzetti}).

\subsection{Features}

\begin{figure}
\begin{tabular}{c}
\includegraphics[width=25pc]{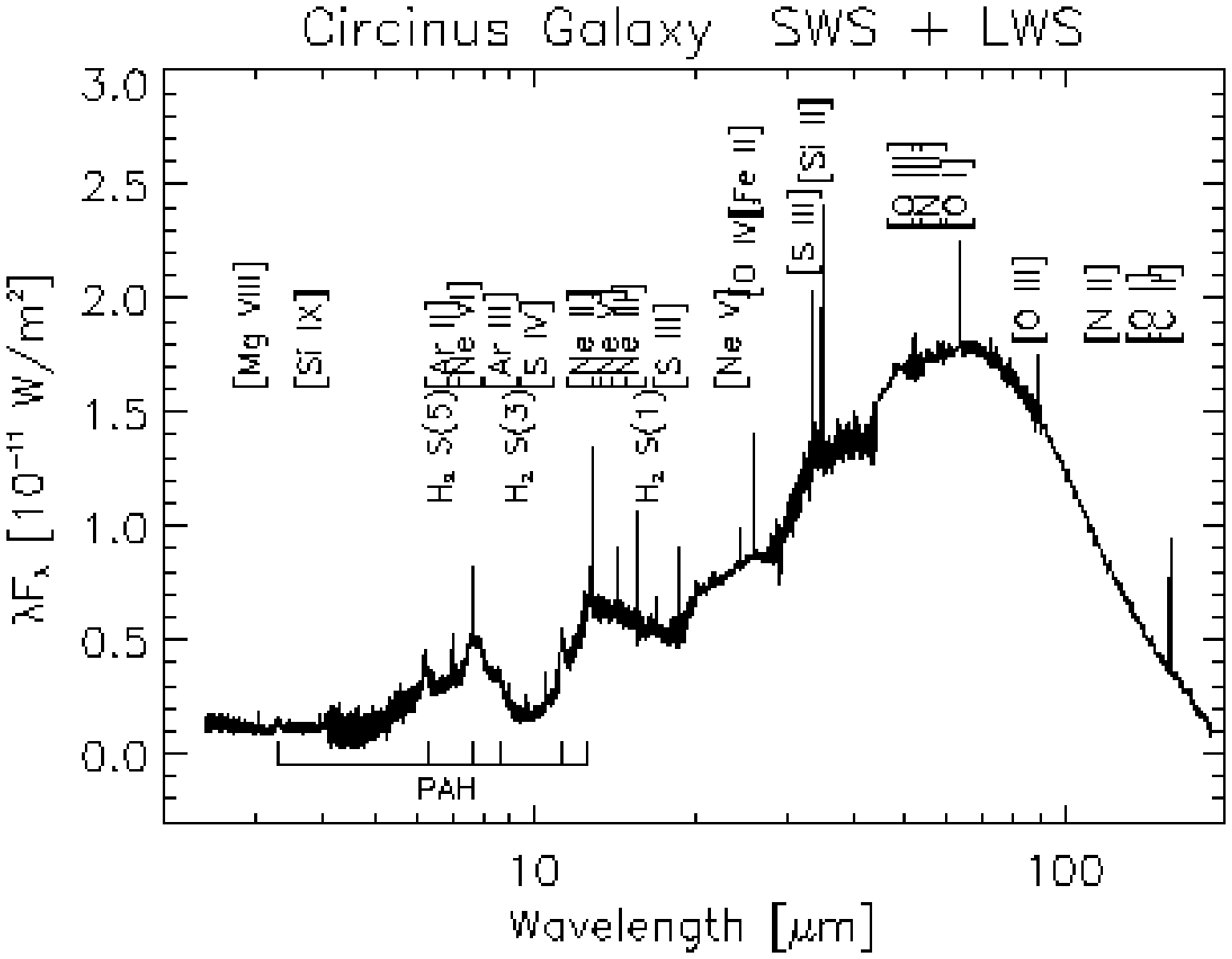}\\
\includegraphics[width=24pc]{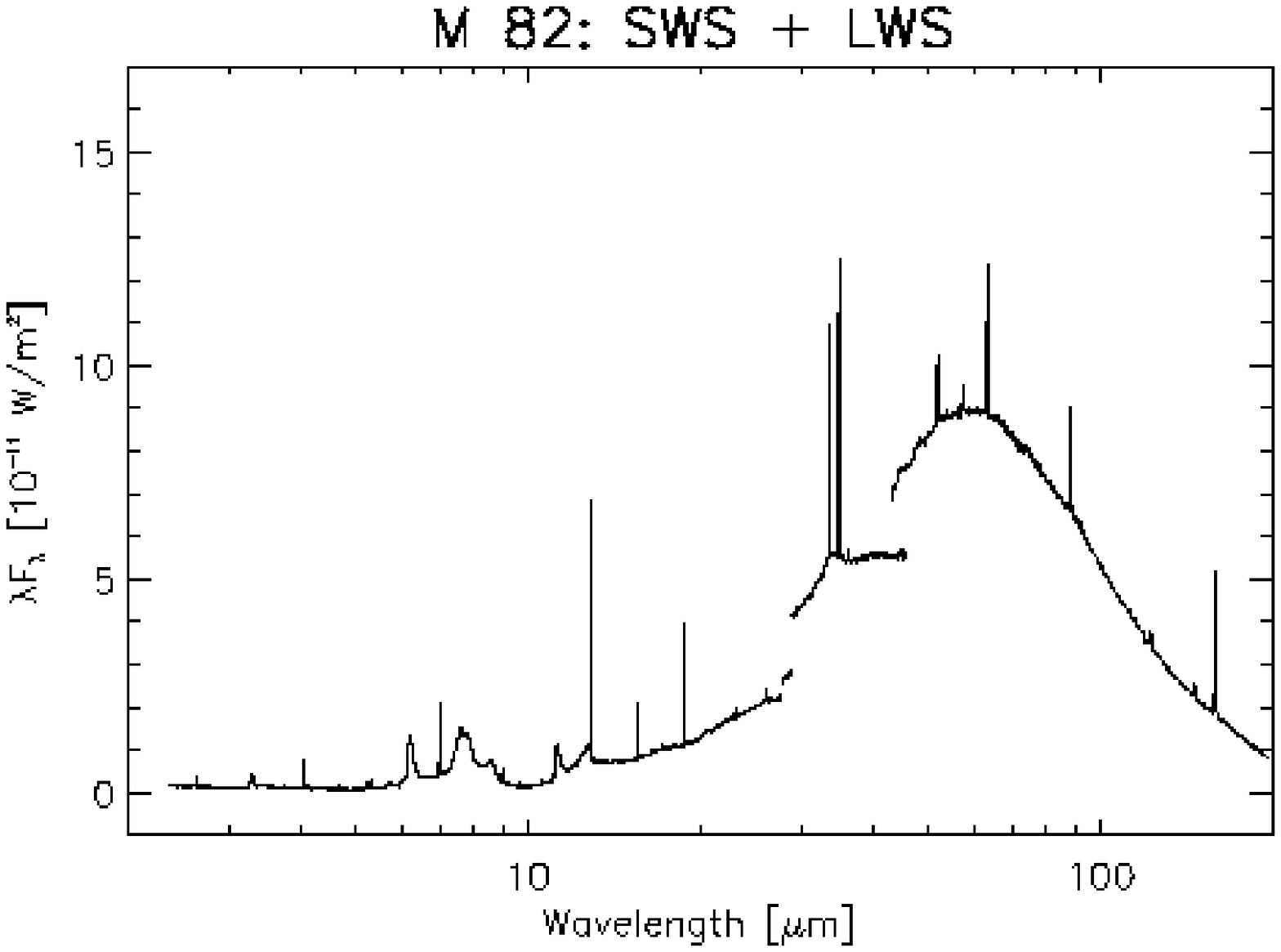}\\
\includegraphics[width=24pc]{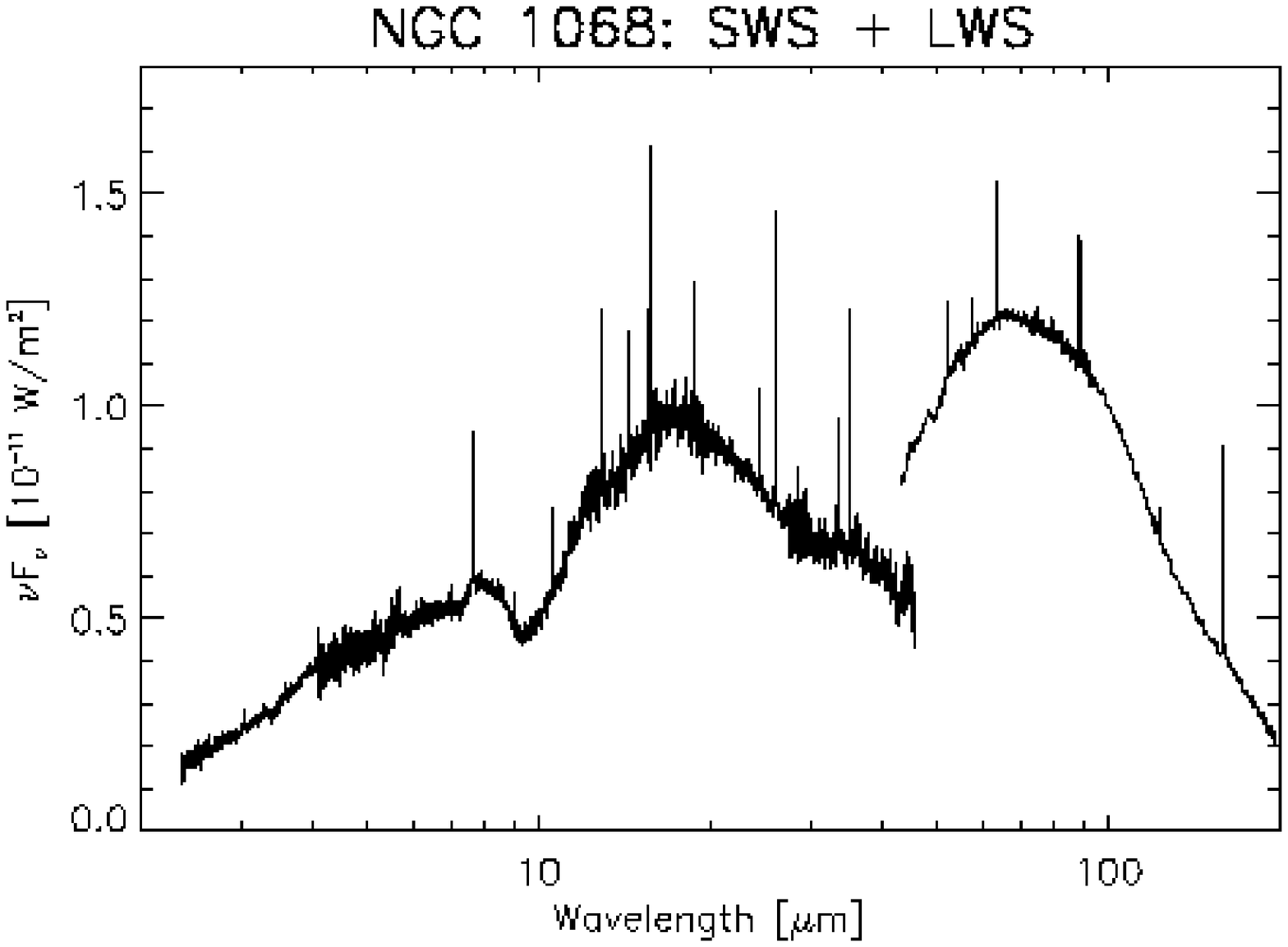}\\
\end{tabular}
\caption{Composite SWS+LWS spectra of Circinus, M82 and \n1068 displaying the spectral differences between the different source types (Sturm priv. comm.).}
\label{circinus}
\end{figure}

The advent of \iso\ spectroscopy enabled 
the first sensitive and high resolution
mid- to far-\ir\ spectroscopic measurements of active galaxies providing a better
understanding of the physical conditions within visually obscured
regions of systems such as Circinus \cite{1996A&A...315L.109Moorwood,2000A&A...358..481Sturm}, the starburst M82 \cite{2000A&A...358..481Sturm}
and the Seyfert 2 NGC1068 \cite{2000ApJ...536..697Lutz}. The full
2.5-200\mn\ spectra of these three galaxies are shown in Fig. \ref{circinus},
and display the richness of the \ir\ range and the differences between these active galaxy templates \cite{2000A&A...358..481Sturm}.
The composite source Circinus displays broad
\mir\ features, fine structure lines of low excitation, \hrl s and a
thermal continuum that predominantly trace the starburst component are intermingled
with the strong VSG and \mir\ AGN continua and high excitation \fsl s
that are predominantly caused by the hard radiation field of an
AGN. In addition, warm and cold molecular gas and warm atomic gas are
probed by rotational lines of H$_2$, molecular absorption features
(e.g. OH, CH, XCN, and H$_2$O) and very low excitation (<13.6eV) \fsl
s (e.g. [OI], [CII]), respectively.

\subsubsection{Unidentified Infrared Bands}
The \mir\ spectra of many active galaxies are dominated
by characteristic broad emission features, the most prominent of which are located
at 3.3, 6.2, 7.7, 8.6 and 11.3\mn.
They are invariably observed
together and are commonly referred to as unidentified infrared bands
(UIBs). These features have been detected in a wide range of celestial sources
(e.g \opencite{1995fgts.symp...23Allamandola}; \opencite{1996A&A...315L.309Cesarsky}; \opencite{1996A&A...315L.305Cesarsky}; \opencite{1996A&A...315L.337Verstraete}; \opencite{1996A&A...315L.353Mattila}; \opencite{1998A&A...331..742Lemke}; \opencite{1999A&A...342..643Mattila}) and are commonly detected in the spectra of normal galaxies.
UIBs may originate from C-C and C-H bending and stretching vibrational modes in aromatic
hydrocarbons
\cite{1984A&A...137L...5Leger,1985ApJ...290L..25Allamandola,1989ARA&A..27..161Puget} and the favoured carriers are polycyclic aromatic hydrocarbons (PAHs). The
features are attributed to PAH clusters \cite{1998A&A...339..194Boulanger}, while individual PAHs can contribute to the 5-10\mn\ continuum
and may be responsible for the
non-linear continuum rise seen in the far UV
\cite{1992A&A...259..614Siebenmorgen}. Alternative carriers are small amorphous carbon grains that emit when exposed to moderately
intense UV/visible radiation
\cite{1989ARA&A..27..161Puget,1996A&A...315L..85Boulade,1998ApJ...493L.109Uchida,1999A&A...351..447Pagani}.
Since the exact molecular structure of the carriers remains under
debate, the features are referred to as UIBs throughout this text (for
a more thorough review on the structure of UIBs see the chapter of
Peeters et al. in this volume or
\inlinecite{2003astro.ph.12184Peeters}). Where strongly detected, the
carriers are likely to provide the majority of the flux in the
3-12\mn\ band and show near-invariant profiles in a variety of
galaxies
\cite{1999AJ....118.2625Rigopoulou,2000ApJ...532L..21Helou,2003A&A...399..833Foerster}. The widths of the lines are thought to arise from temperature broadening of the vibrational bands and have been fit using Lorentzian rather than Gaussian profiles \cite{1998A&A...339..194Boulanger,2001A&A...372..981Verstraete} but see the discussion in Peeters et al. (this volume) for further details on this issue.
The spatial correlation between \mir\ emission with atomic and
molecular hydrogen suggests that UIB emission originates at the
interface of \hii\ and molecular clouds, the so-called photo-dissociation regions
(PDRs, e.g. \opencite{1996A&A...315L.337Verstraete};
\opencite{1996A&A...315L.305Cesarsky};
\opencite{1999A&A...351..447Pagani}).

Mid-infrared starburst spectra are generally dominated by these UIB features
and show a strong continuum from
very small grains.
The relative strengths of the features are sensitive to local
radiation fields, distribution of grains throughout the galaxy system
and extinction
\cite{1996A&A...315L.337Verstraete,2000A&A...358..481Sturm,1998ApJ...505L.103Lutz,2000A&A...359..887Laurent}.
UIB emission in starburst regions originates from UV-heated ionised
UIB carriers. A significant contribution can also arise in heating of
neutral UIB carriers in the extended regions by the diffuse ISRF
i.e. by less energetic (visible) photons, which is mostly seen in
normal galaxies and is proposed in some active galaxies \cite{1999A&A...342..643Mattila,1998ApJ...500..181Smith,1999A&A...351..447Pagani,2002A&A...385L..23Haas}.

In intense radiation fields ($\sim$10$^5$ times that of the local
ISRF) such as in metal poor environments or in the
vicinity of super star clusters or AGN, UIBs are weak or absent
\cite{2000A&A...362..310Contursi,1999ApJ...516..783Thuan,1999AJ....118.2625Rigopoulou,2000A&A...358..481Sturm,2001A&A...369...57Santos}. Small
grains and UIB carriers are susceptible to high temperature
fluctuations and photons with sufficient energy may cause grain
sublimation or destruction. This is reflected in the UIB-to-\mir\
continuum strength ratio (7\mn/15\mn\ flux density) that is seen to be lower ($>$0.5) in active regions and galaxies
than in more quiescent regions such as in normal galaxies and the disks of some active galaxies [ratio$\sim$1 with a large dispersion, 
see Fig. 1 of
\inlinecite{1999usis.conf..805Vigroux} and
Fig. 12 in \inlinecite{2000AJ....120..583Dale}]. 
Weak detections of UIBs in galaxies with AGN probably originate from the
cooler material such as
circumnuclear starburst regions or diffusely heated extended regions
rather than from the immediate vicinity of the AGN torus. 
In nearby Seyferts, spatially
resolved \mir\ spectroscopy 
suggested that the absence or suppression of UIB emission is due to the fact that the dust is predominantly heated by processes related to the central AGN
[e.g. in \n1068
\cite{2001A&A...367..487Lefloch}, Circinus
\cite{1999usis.conf..825Moorwood}, \n4151
\cite{1999ApJ...512..197Sturm}, Mrk 279
\cite{2001A&A...369...57Santos}].
This was confirmed through a comparison of \iso\ and high spatial
resolution ground based data of AGN and starbursts, that demonstrated the
absence of UIBs in the nuclei of AGN-hosting galaxies
\cite{2004A&A...414..123Siebenmorgen}. For unresolved sources the VSG continuum may be so strong that the UIB features are diluted by the over-powering continuum resulting in their non-detection \cite{1998ApJ...505L.103Lutz}. 
Thus at high redshifts, where low-metallicity- and/or AGN-dominated- systems may
be more prevalent, the intrinsic weakness of UIB features seen in
local systems of these types would imply that the use of UIB features as a tracer of similar systems at high redshift may be problematic.

\subsubsection{Fine structure lines}
\label{fsl}
Fine structure lines are tracers of nebular conditions such as
excitation and the intrinsic ionising spectrum. By virtue of their differing excitation potentials and critical densities 
they provide an insight into the energetics and chemical composition of the regions %
from which
they originate. 
Starburst galaxies commonly show low excitation \fsl
s (e.g. [ArII], [ArIII], [FeII], [NeII] and [SIII]) that have ionisation
potentials between 13.6eV and $\sim$50eV and mostly arise from
\hii\ regions that have been photoionized by their central massive
stars. A PDR or shock
 origin may also contribute to the flux of some low-excitation lines.  
High excitation lines with potentials exceeding 50eV (e.g. [NeV,VI],
[SIV],[MgV,VII,VIII] [OIV], [SiV,VI,VII]) are present in the
spectra 
some low-metallicity galaxies
but are more commonly found in AGN where high energy photons, produced by non-thermal processes related to the central black hole, photoionise the gas.
Excitation to these high energy levels ($>$50eV)
is difficult to attain by photons emitted by stars in typical H\,II
regions. However, the \mir\ spectra of some starbursts do show
the high excitation [OIV] and [SIV] lines. The presence of these lines is
associated to collisionally excited, shocked or
coronal gas (e.g. \inlinecite{1998A&A...333L..75Lutz}). 
Active galaxy spectra also show \fsl s with excitation potentials
below the ionisation potential of hydrogen (e.g. [OIII]52,88\mn,
[OI]63,145\mn\ and [CII]158\mn), that trace mostly cool atomic and
molecular clouds.
[CII]158\mn\ and [OI]63\mn\ 
are important cooling lines in PDRs but
also originate in widespread atomic gas (see section \ref{cii}).

As infrared fine structure lines are fairly insensitive
to uncertainties of the effective temperature of the gas, and because they are
less affected by extinction than optical lines, they
provide an
unequalled insight into the properties of the ISM
(e.g. electron
temperature and density, ISRF, metallicity) of active galaxies and
their embedded power source. Ratios of two lines of the same species
provide information on the nebular conditions such as electron
temperatures and densities. Ratios of high to low excitation states of
the same element probe the hardness of the ISRF and ratios of high- to
low-excitation lines of different elements have been used as tracers
of AGN presence and thus contribute to nebular diagnostics (see
section
\ref{diagnostics}). Spectroscopic data from \iso\ was a major advance over \iras\ 
enabling the quantitative estimation of the dominant fuelling mechanism
in a given source
\cite{1998ApJ...498..579Genzel,2000A&A...359..887Laurent,2000A&A...358..481Sturm}. 

\subsubsection{Molecular Features}
Molecular features trace the cold dense molecular material in active galaxies.
Several active gas-rich starbursts, Seyferts and \uls\
display a range of molecular lines: the rotational lines of molecular hydrogen (Sect. \ref{h2}),
OH, H$_2$O, NH$_3$ and CH are seen in absorption or emission (see Fig. 1 of \opencite{1999Ap&SS.266...91Fischer}, also refer to
\opencite{1996AAS...189.8904Hur}; \opencite{1997AAS...191.8903Fischer}; \opencite{1999usis.conf..861Bradford}; \opencite{1999usis.conf..817Fischer}; \opencite{1999Ap&SS.266...91Fischer}; \opencite{1999ApJ...511..721Colbert}; \opencite{2000A&A...358..481Sturm}; \opencite{2002A&A...389..374Rigopoulou}; \opencite{2002A&A...385.1022Spoon}; \opencite{2005ApJ...623..123Spinoglio}). 
Molecular absorption features appear to be more common (and prominent) in sources of high luminosity (e.g. \uls) than those of lower luminosity 
\cite{1999Ap&SS.266...91Fischer,1999usis.conf..817Fischer,2002A&A...385.1022Spoon}.  
Spectra displaying molecular absorption features show a wide variety of
excitation: from starburst galaxies like M82 which display features indicating
that most of the O and C molecules occupy the ground-state, to objects such
as \uls\ where a significant population exists at higher energy levels
\cite{1999Ap&SS.266...91Fischer,1999usis.conf..817Fischer}. 
Additional absorption features at 6.85 and 7.25\mn\ are seen in the
spectra of some active galaxies and \uls\
\cite{2002A&A...385.1022Spoon}.
They are attributed to CH-deformation modes of carbonaceous
material because of their similarity to features seen along lines of sight towards Sgr A* 
\cite{2000ApJ...537..749Chiar}. \ul\ and Seyfert spectra also exhibit an
absorption feature at 3.4\mn\ 
that is likely to be aliphatic CH stretch absorptions of refractory carbonaceous material \cite{2002ApJ...569...44Imanishi,2004A&A...423..549Dartois}.

\paragraph{Silicate Absorption}
The 9.7\mn\ and 18\mn\ absorption features arise from the
stretching and bending modes of SiO. The extinction suffered by a
galaxy can be estimated by the depth of these lines. 
For example, the extreme 9.7\mn\ silicate absorption feature seen in
the starburst-\ul\ Arp 220 provides a lower limit to the optical depth
of A$_{V} \gsim$45 (\inlinecite{1997eai..proc..283Charmandaris}, but
see \inlinecite{2004A&A...414..873Spoon} for a more detailed analysis
of this system).  Though it is possible that the depth of the 9.7\mn\
may be overestimated in the feature-dense
\mir\ spectra of starbursts exhibiting UIB emission,
\cite{2000A&A...358..481Sturm,2003A&A...399..833Foerster}.
The extreme extinction of the starburst M82 (A${_V}=$15-60) derived by
\inlinecite{1975ApJ...198L..65Gillett} was ascribed to this problem as \iso\
data could not support such a high extinction
\cite{2000A&A...358..481Sturm}.  

\paragraph{Molecular hydrogen and warm molecular gas}
\label{h2}
Rotational transitions of H$_2$ that originate in dense molecular
media were detected in the spectra of several active galaxies: \n3256
\cite{1996A&A...315L.125Rigopoulou}; \n891
\cite{1999ApJ...522L..29Valentijn};
\n1068 \cite{2000ApJ...536..697Lutz}; \n4945
\cite{2000A&A...357..898Spoon}; \n6240
\cite{2003A&A...409..867Lutz}. Low rotational transitions
originate from warm gas (T$\sim$100-200K) and the higher 
and rovibrational lines from much hotter gas
(T$\gsim$1000K). \inlinecite{2002A&A...389..374Rigopoulou}
investigated
a sample of starbursts and Seyferts observed by \iso\ presenting
pure rotational lines from S(7) to S(0).
Irrespective of starburst or Seyfert type,
temperatures of $\sim$150K are derived from the S(1)/S(0) ratio. A similar temperature was found for \n4945
\cite{2000A&A...357..898Spoon}. While $\sim$10\% of the galactic gas mass
in a starburst system can be accounted for by the warm component, on average
higher fraction (18\%, range 2-35\%) are observed for Seyferts \cite{2002A&A...389..374Rigopoulou}. The temperature of the molecular hydrogen implies that it arises from
a combination of emission from fairly 
normal PDRs (at least for starbursts), 
low velocity shocks and X-ray heated gas around the central AGN. The latter is
more probable in Seyfert galaxies which accounts for the larger fraction of warm
gas found in these systems. Shock heating appears to be the most likely origin of the
extremely strong rotational H$_2$ lines detected in the merging
'double active nucleus' system \n6240 \cite{2003A&A...409..867Lutz}.
While, the origin of shock-heating can be largely associated to winds
originating from massive stars and supernovae, recently
\inlinecite{2005A&A...433L..17Haas} have found the first observational
evidence of pre-starburst shocks in the overlap region of the early
merger the Antennae (see Section \ref{mergers}).  From a re-analysis of
archival \cam-CVF data, they find that the strongest molecular
hydrogen emission (traced by the H$_2$S(3) line) is displaced from the
regions of active star foramtion. This, together with the high line
luminosity (normalised to the \fir\ luminosity) indicates that the
bulk of excited H$_2$ gas is shocked by the collision itself.

\paragraph{Ice absorption and cold molecular gas}
\label{ices}
An interesting discovery made by \iso\ was the first extragalactic
detection of absorption features due to ices (H$_2$O, CH$_{4}$ and
XCN) present in cold molecular components of starbursts, Seyferts and
predominantly \uls. \inlinecite{2000A&A...358..481Sturm} first
reported the detections of weak water ice absorption features at 3\mn\
in the spectra of local starbursts M\,82 and \n253.  Very strong water
ice, as well as the first detections of absorptions due to CO and
CO$_{2}$ ices, were detected in the actively star-forming galaxy
\n4945 \cite{2000A&A...357..898Spoon}.
The presence of ices is related to the presence of cold material and
the radiation environment. One would expect cold molecular clouds in
starbursts to be likely hosts while more intense environments, such as
those predominantly fuelled by AGN, to be less favoured. 
This is corroborated by the lack of absorption features
in the spectrum of \n1068
\cite{2000A&A...358..481Sturm} and the limits on the strength of the CO absorption feature set from \sws\ spectra of 31 AGN \cite{2004A&A...426L...5Lutz}.
However, water ice absorption has been detected in Seyfert galaxies,
e.g. in the UIB-free spectrum of the Seyfert galaxy \n4418
\cite{2001A&A...365L.353Spoon}. The absorption dominated spectrum
bears strong similarities to the spectra of embedded protostars, and
the depth of absorption features implies that its Seyfert nucleus is so deeply embedded that ices are shielded from
the intense radiation field of the AGN. 

In a heterogeneous sample of 103 active galaxies with high
signal-to-noise \mir\ spectra, approximately 20\% display absorption
features attributed to ices \cite{2002A&A...385.1022Spoon}.
While ices are weak or
absent in the spectra of starbursts and Seyferts, in \uls\ they are
strong, considerably stronger than the weak features found previously
in M\,82 and \n253. This implies that rather than the radiation
environment it is the amount of cold material that determines their
presence in these most luminous and massive infrared galaxies that
are known to contain large concentrations of molecular material
(e.g. \inlinecite{1997ApJ...478..144Solomon}; \inlinecite{1998ApJ...507..615Downes}).
Although plausible,
more sensitive (and spatially resolved) \mir\ spectroscopy is required
to understand these differences, and the distribution and energetic
conditions of cold molecular material. Moreover, higher
spectral resolution over a wider \mir\ wavelength range is necessary
for accurate decomposition of the spectra into their various
components, which will aid identification of subtle and/or blended
features (e.g. \opencite{2004A&A...414..873Spoon}). 
\inlinecite{2002A&A...385.1022Spoon} proposed an evolutionary sequence
for their ice spectra ranging from a highly obscured beginning of star
formation where ices dominate the spectra, to a less obscured stage of
star formation where the UIBs become stronger. At any stage the star
formation may be accompanied by AGN activity.
\paragraph{OH features and (Mega-)masers}

Transitions of the OH molecule were also detected in the \sws\ and \lws\ spectra
of some active galaxies \cite{1999Ap&SS.266...91Fischer}, such as in the starbursts \n253
\cite{1999usis.conf..861Bradford} and \n4945
\cite{1997AAS...191.2105Brauher} permitting estimation of the OH
column density and abundance.  The depths of the OH features detected
in the \lws\ \fir\ spectra of galaxies hosting known OH (mega-)masers
were found to be sufficient to provide all of the \ir\ photons
required to radiatively pump the maser activity (e.g. in the starburst-\ul\ Arp 220
\cite{1997Natur.386..472Skinner,1998AAS...193.9002Suter}; AGN-\ul\ Mrk 231
\cite{1998AAS...193.9002Suter}; the merging double systems \ul\ IRAS 20100-4156 and \lg\ 3 Zw 35
\cite{1999A&A...351..472Kegel}; the starburst \n4945
\cite{1997AAS...191.2105Brauher}).
In Arp220, the fact that the \fir\ continuum source is extended and
that the maser must be located in front of it
favours the interpretation that a starburst provides the continuum
photons for its maser \cite{1997Natur.386..472Skinner}. The high pump
rate determined for this source suggests that pumping mechanisms other
than radiative pumping (such as collisional pumping) may contribute
\cite{2004NewA....9..545He}.
\subsection{Magnetically Aligned Dust Grains}

Polarised emission detected from active galaxies is ascribed to dichroic
absorption by the ISM of a galaxy i.e. by elongated dust grains
that are magnetically aligned. This has been observed in galactic
sources where the position angle of the measured polarisation
indicates the orientation of the projected magnetic field.
Polarisation is an important probe of the physical conditions in active galaxies
as it can
provide constraints on the formation of the putative torus
\cite{1999ipo..work....1Alexander}. Such 
polarisation studies were performed for some \uls\ and revealed 3-8\% of the 5-18\mn\ flux
to be polarised \cite{2001A&A...376L..35Siebenmorgen}. The archetypal
starburst-like \ul\ Arp220 has the lowest polarisation while the warm
AGN-like Mrk 231 \ul\ the highest. The first polarisation profile of a
starburst galaxy was made at 6\mn\ for \n1808 showing a 20\% increase towards the
outer regions \cite{2001A&A...377..735Siebenmorgen}. 
This, together with a complementary
measurement at 170\mn\ (unresolved), are best explained by the large scale
($\sim$500pc) magnetic alignment of (non-spherical) large grains ($\ge
100$\AA). 
\section{Very Cold dust and diffuse radiation fields: ISM}
\label{verycold}
\pht-data clearly confirmed the presence of very cold dust
(T$\sim$10-20K) in a variety of infrared galaxies.
Often in
conjunction with supplementary (sub-)mm photometry, this component was
detected in spatially resolved maps or intensity profiles of local
galaxies (e.g. \inlinecite{1998A&A...338L..33Haas}), and was isolated
as a significant contributor to the \fir\ luminosity in SED analyses
of unresolved active galaxies \cite{2000ApJ...533..682Calzetti,2001A&A...379..823Klaas,2002ApJ...572..105Spinoglio,2003A&A...402...87Haas}. 
Although cold 'cirrus' or 'disk' components in galaxies were indicated by \iras\ colours, \iras\ was insensitive to this very cold
component due to its shorter wavelength coverage ($<100$\mn). As a result 
gas-to-dust ratios of distant galaxies based solely upon \iras\ data were 
overestimated whereas the addition of \iso\ measurements 
resulted in values which are 
closer to that of the Galactic value ($\sim$160) \cite{1998A&A...338L..33Haas}
The very-cold component often accounts for a
substantial fraction of \fir\ emission and is most commonly seen in
early type and disk galaxies (see Sauvage et al. this volume). With a scale-length in excess of the stars, less than the atomic HI gas and 
similar to that of the molecular gas, it is
likely that the dust is associated to this cold gas components
(e.g. \opencite{1999MNRAS.304..495Davies};
\opencite{2003A&A...410L..21Popescu}; \opencite{2004A&A...414...45Popescu}).

Very-cold dust is also seen in both starburst and
AGN-hosting active galaxies. \n253,
that hosts a nuclear starburst, has been shown to contain extended
cold dust along the major and minor axes. While the dust along the
minor axis is attributed to outflows, the cold emission along the
major axis is shown to have a scale length 40\% greater than that of
the stars
\cite{2001A&A...377...73Radovich}. 
The correlation of the 90\mn\ emission profile of four Seyfert
galaxies with the R-band extent but not with the 
star formation tracer H${\alpha}$ suggests that a
large contribution of the \fir\ emission from Seyferts does not arise
from the (nuclear) star-forming regions but from a cooler \fir\ disk
component within which normal stars heat dust grains
\cite{2000ApJ...529..875Perez}. The detection of substantial cold \fir\ emission
(T$<$20 K) from the star-forming obscured overlap region of the
merging Antennae galaxy (see Section \ref{mergers}) indicates the
presence of cold dense dust within which stars are forming
\cite{2000A&A...356L..83Haas}.

SED analysis using black or grey-body decomposition
suggested the need for more than one dust component, including a very
cold dust component, to explain the emission above 100\mn\ of many galaxies.
However, by allowing both temperature and dust grain emissivity to be free parameters, it is possible to approximate 
the entire \fir\ SED ($\lambda>$50\mn) using a
a single modified black body.
Invariably for active galaxies this was achieved with temperatures
$\sim$30-50K and emissivity values of 1.5-2 %
(e.g. \opencite{1999ApJ...511..721Colbert};
\opencite{1999A&A...351..495Siebenmorgen}).  However, by doing so raised conflicts with other observational constraints 
(see \inlinecite{2001A&A...379..823Klaas} for
a full description). Moreover by leaving the dust emissivity as a free
parameter presupposes that intrinsic dust properties differ from
source to source with no physical basis supporting this assumption. By
choosing a Galactic value of $\gamma$=2 for the grain emissivity
(i.e. assuming that dust properties do not vary between sources),
single black-bodies fail to explain all the \fir\ emission -
additional cold components (T$\sim$10-20K) are required that are
associated with the spatially resolved extended very cold components
described above.

The SEDs of active galaxies are warmer than those of normal galaxies
and the \fir\ part is generally fit by two components representing 
active star formation and very cold (diffusely heated) material. The
very cold dust can be a significant contributor in terms of galaxy
mass and energy output. For example, two of the three components used to fit the SED of the LINER \n3079 have very cold temperatures of 12
and 20K, respectively \cite{2002A&A...391..911Klaas}.
\inlinecite{2000ApJ...533..682Calzetti} find that two dust components
are required at T=40-55K and T$\sim$20-23K to explain the \fir\ SEDs
of eight local starbursts. The latter comprises up to 60\% 
of the total flux and the associated mass is as much as
150 times that of the former. 
For the most actively star forming galaxies or those with strongest
radiation fields, the contribution to the bolometric luminosity from
cold emission (as measured in the 122-1100\mn\ range) was shown to be
smaller in actively star forming (5\%) than in cirrus dominated
galaxies ($\sim$40\%) \cite{2001ApJ...549..215Dale}. SED analysis
of a wide range of
active galaxies indicates that very cold components are common whether
they are centrally fuelled by AGN or starbursts, suggesting that the central activity does not directly affect the \fir\ emission.
\section{Cold dust and low excitation radiation fields: starburst regions}
\label{cold}

The T=40-45K temperature component found in local starbursts by
\inlinecite{2000ApJ...533..682Calzetti} mentioned
above represents the emission from dust surrounding sites of active
star formation. For starburst galaxies this emission (peaking between 60 and 100\mn) dominates
the infrared SED and is the major contributor to the infrared
luminosity. This 'cold' \ir\ emission is generally less extended than the very cold component described in the preceding section.
In some nearby galaxies, the \fir\
morphology traces star forming regions well. For example, the
strongly \fir-emitting dark molecular cloud coincides with a cluster
of star forming regions in the galaxy \n4051
\cite{2000ApJ...529..875Perez}. Similarly, \fir\ emission could be
resolved for the pair members of
interacting spirals where enhanced star formation has been established
\cite{2000ApJ...541..644Xu,2001AAS...198.3603Gao}. 
The correlation of the warmer \fir\ emission with the
R-band morphology of local Seyferts hosting compact nuclei and
circumnuclear star formation further supports that this 
\fir\ component is related to star formation activity
\cite{1997ApJ...487L..33Rodriguez}. 
Moreover, \fir\ emission has also been spatially associated with
regions exhibiting signs of strong star formation (e.g. strong
UIBs).  \inlinecite{2001A&A...372..427Roussel} demonstrated that the integrated \fir\ flux and the
star-formation related \mir\ flux are strongly correlated in the disks of 
normal star forming galaxies indicating that the grains responsible for the
mid- and \fir\ share a common heating source. 
\subsection{The importance of interactions/mergers}
\label{mergers}
\begin{figure}
\centerline{\includegraphics[width=25pc]{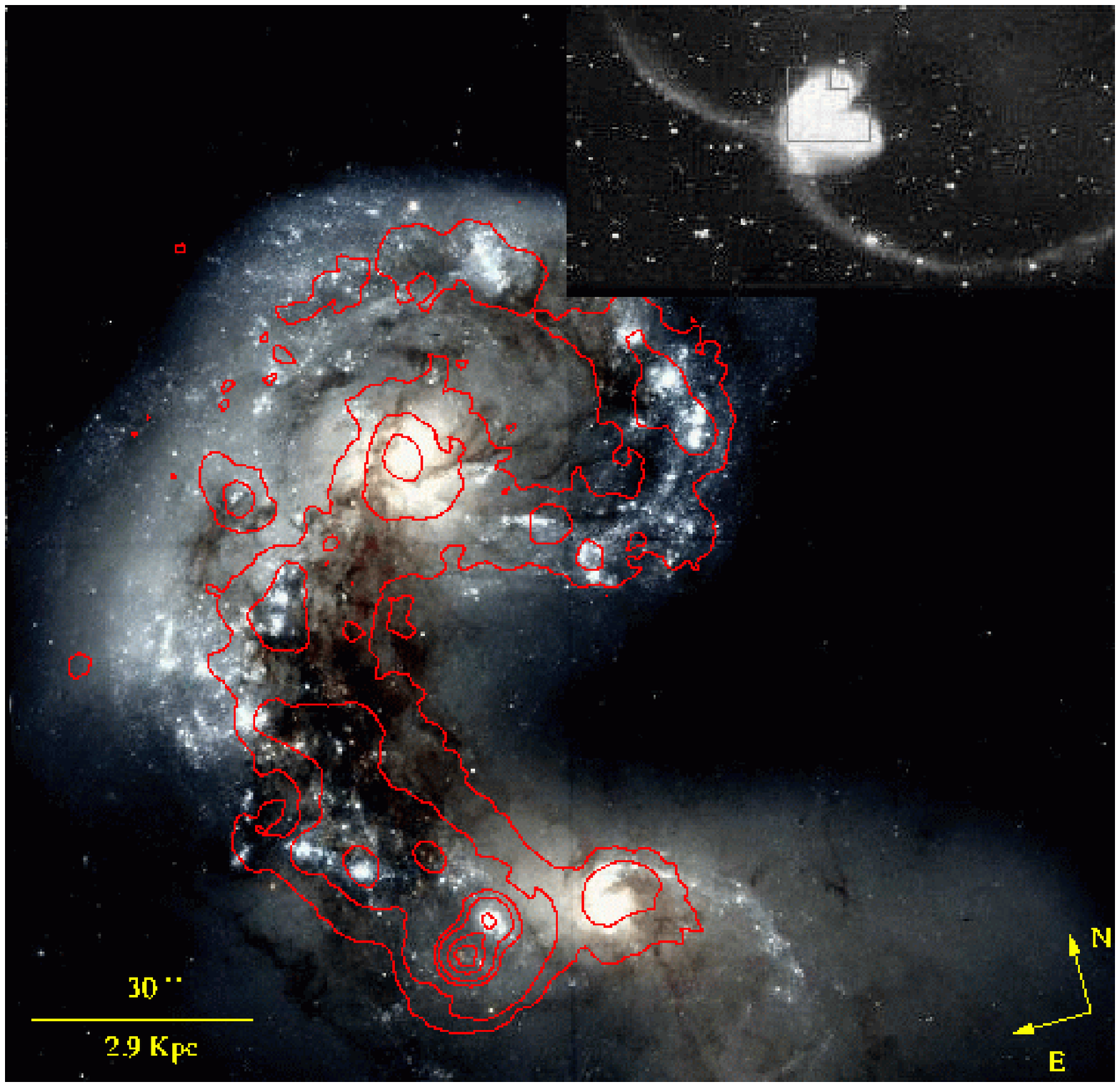}}
\centerline{\includegraphics[width=15pc]{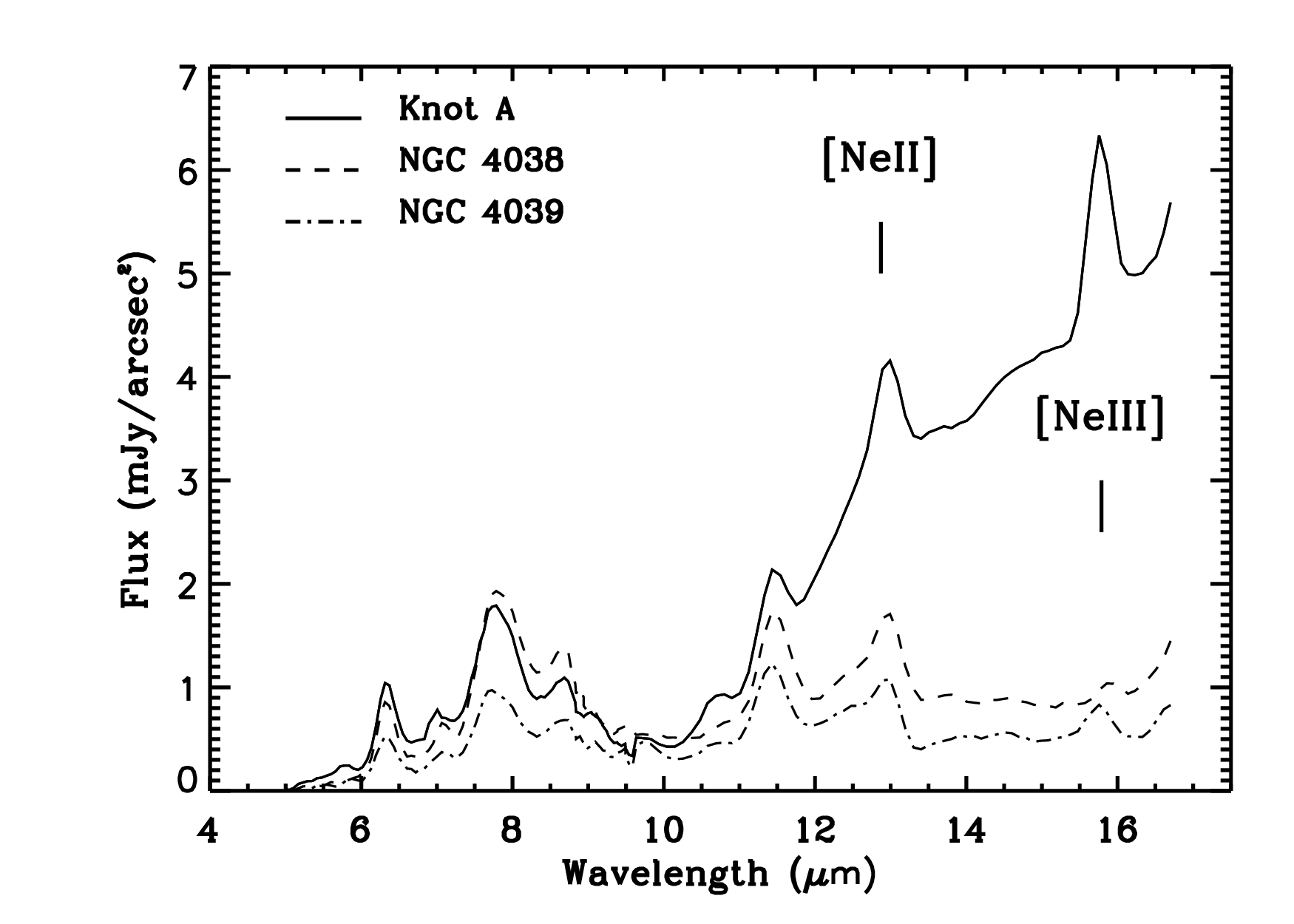}}
%\centerline{\includegraphics[width=25pc]{VermaA_2a.pdf}}
%\centerline{\includegraphics[width=15pc]{VermaA_2b.pdf}}
%
\caption{This composite figure of the Antennae is from Mirabel et al. (1998) showing the \cam-CVF contours overlaid on a HST V and I band image. Half of the \mir\ emission arises from starburst activity that is completely obscured in the optical, including the brightest \mir\ knot (knot A). The CVF spectrum of this knot is shown in the lower part of the image displaying strong signatures of massive star formation - the NeIII line and a strong rising continuum above 10\mn.}%
\label{antennaefig}
\end{figure}

Interactions and mergers enhance starburst activity, both nuclear and
extra-nuclear (see review by \opencite{1999PhR...321....1Struck}).%
The final merger stage plays a critical role in how active galaxies form stars and
may influence the appearance of AGN and the onset of the \ul\
phase. The fraction of disturbed or interacting systems in infrared
selected samples of galaxies appears to increase with luminosity with
nearly 100\% of all \uls\ displaying evidence of interaction (see
\opencite{1996ARA&A..34..749Sanders} and references therein).  
Numerical simulations have also helped in establishing that 
interactions and mergers play an
important role in the formation and evolution of such galaxies
\cite{1996ApJ...464..641Mihos}.
The induced gravitational instabilities form bars
which strip in-falling gas of angular momentum and enable radial
inflows to the nuclear regions that can feed starbursts or AGN
\cite{2001ApSSS.277...29Combes}. Moreover, mergers appear to be the
triggering mechanism for the ultraluminous phase. Dynamical shocks and
tidal forces are the only mechanisms that can cause transference and
compression of sufficient quantities of matter to fuel luminosities in
excess of $10^{11}L_{\odot}$. 

\cam 's spatial scale was sufficient to
map the morphologies of local interacting active systems (see
\opencite{1999Ap&SS.269..349Mirabel} for a review) and agree with
results from ground-based high resolution images that (optically-faint
or invisible) compact sources located in deeply obscured nuclear
regions or interaction interfaces dominate the \mir\ emission
(e.g. \opencite{2000AJ....119..509Soifer}; \opencite{2001AJ....122.1213Soifer}; \opencite{2002A&A...391..429Charmandaris};
\opencite{2004A&A...414..845Gallais};
\opencite{2004A&A...414..123Siebenmorgen}). 
For some sources, diffuse
emission between these 'hot-spots' can also contribute significantly
to the total \mir\ emission
(e.g. \opencite{2002A&A...391..417Lefloch};
\opencite{2004A&A...414..845Gallais}). This diffuse emission can be
enhanced in merging systems that are common among active galaxies.
By probing through the famous
characteristic dust lane of the AGN-hosting elliptical galaxy
Centaurus A (NGC 5128), imaging with \cam\ revealed a barred 'mini-spiral'
located in its central 5 kpc.  The mini-spiral was proposed to be
tidal debris from a gas rich object that was accreted in the last
gigayear
\cite{1999A&A...341..667Mirabel,2000A&A...353...72Block}. 
Evidence of accreted companions in Cen A exists from previous studies of its stellar and gas kinematics
(\opencite{2000A&A...356L...1Charmandaris}, and references therein). \pht\ contributed to the accretion scenario through the 
detection of the
northern cold dust shell that is possibly the ISM
remnant of a captured disk galaxy
\cite{2004A&A...415...95Stickel}. As a result Cen A is a prime example of a giant elliptical that must have experienced several minor mergers. 
It is likely that, as the interactions took place, tidal forces
funnelled gas into the dynamical centre and boosted the star formation
rates in the gas rich components forming the dust observed in the \ir\
and visible.

\subsubsection{Nuclear and extra-nuclear starbursts}
Disturbed morphological signatures such as rings, tidal tails, bridges
and arcs are clearly detected in \cam\ observations of numerous merging
galaxies (e.g. the Antennae \cite{1998A&A...333L...1Mirabel}
Fig. \ref{antennaefig}; UGC12915,12914 \cite{1999AJ....118.2132Jarrett};
\n985 \cite{2002ApJ...566..682Appleton}; Arp 299 \cite{2004A&A...414..845Gallais}). The \mir\ morphologies depend upon the mass
ratio of the progenitor galaxies and the merger stage. As well as
enhancing star formation in the nuclear regions, interactions cause a
general increase in the widespread activity of the parent galaxies.
At the early stages of interaction, wide scale star formation can be
induced and dust grains are heated by a generally warmer radiation
field. For example, \cam\ observations of the VV 114 galaxy show that
60\% of the \mir\ flux arises from a diffuse component that is several
kpc in size \cite{2002A&A...391..417Lefloch}. Nearly forty percent of the 7
and 15\mn\ emission detected in the early-stage merger
(nuclei separation
5kpc) galaxy pair Arp 299 (\n3690 and IC 694) is
diffuse and originates from the interacting disks
\cite{2004A&A...414..845Gallais}.
In more advanced mergers, shocks and gravitational instabilities
 encourage the onset of star formation. The transfer of matter to
 nuclear regions and compression in interaction zones gives rise to
 enhanced \mir\ emission in nuclear and often extra-nuclear sites. A
 clear example of extra-nuclear star formation was uncovered by \cam\
 in the intra-group medium of Stephan's Quintet
 \cite{1999ApJ...512..178Xu}.

The best example of \cam's capability in probing through highly extincted regions
is imaging of the spectacular major merger in the
Antennae (Arp 244, \n4038/4039). The completely obscured overlap
region in the HST-WFPC2 image ($A_V\sim35$mag
\opencite{2003A&A...403..829Verma})
was shown by \cam\ to contain a massive star forming region
($\lsim$50pc in radius) comprising more than 15\% of the total
12-17\mn\ luminosity which outshines the two nuclei by a factor of 5
\cite{1998A&A...333L...1Mirabel}. The warm \mir\ emission 
traces the molecular gas that fuels
star formation \cite{2000ApJ...542..120Wilson,2003ApJ...599.1049Wilson} whereas the cold
($T\sim30K$) and very cold dust ($T\sim20K$) are traced by \fir\ \pht\ and
sub-mm observations \cite{2000A&A...356L..83Haas}.  
\inlinecite{2000A&A...356L..83Haas} find that the sub-millimetre knots, in particular K2, are likely to be in a
pre-starburst phase and the simultaneous presence of powerful off-nuclear starbursts \& 
pre-starbursts may be a general feature of colliding galaxies. 
The authors postulate that once star formation
has commenced in these clouds, the Antennae may evolve from a
luminous IR galaxy into an ultraluminous one.

An interesting feature for the Antennae (also seen for Centaurus A) is
that the most luminous 15\mn\ knots are not coincident with the
darkest, most prominent dust absorption lanes seen in the HST
image. Presumably the darkest optical dust lanes trace the coldest
emission (i.e. that probed by the \fir) rather than the warm dust from which
15\mn\ emission arises. This is plausibly the cause of the
displacement. Alternatively projection effects may be responsible for
this offset
\cite{1998A&A...333L...1Mirabel}.

Extranuclear overlap starbursts emitting in the \mir\ are not unique to the Antennae.
In a sample of 8 spiral-spiral pairs with overlapping disks,
\inlinecite{2000ApJ...541..644Xu} found extranuclear, overlap region
starbursts in 5 galaxies classified as advanced mergers or having
severely disturbed morphologies (but not in less disturbed
systems). However, the starbursts in the
bridges or tidal tails are generally fainter than those in the nuclear
regions.
\subsubsection{Collisional Rings}

Collisional ring galaxies are produced when a
compact galaxy passes through the centre of a larger disk galaxy giving
rise to a radial density wave that creates a massive star forming ring
(see review by
\opencite{1996FCPh...16..111Appleton}). 
If the collision is off centre then the ring morphology is asymmetric
e.g. as in VII Zw 466
\cite{1999ApJ...527..143Appleton}. These systems are excellent 
local examples of induced star formation in colliding galaxies. The effects of collisions on the ISM can be investigated in detail, and may be relevant for higher redshift galaxies including \uls\ \cite{2000isdu.conf..232Appleton}.
Azimuthal variations in over-density in the rings are often noted
as well as extranuclear star forming knots that show strong UIBs
\cite{1999A&A...341...69Charmandaris,1999ApJ...527..143Appleton,2001Ap&SS.276..553Charmandaris,2002ApJ...566..682Appleton}. When detected, star formation
in the ring contributes at a level of $\gsim10\%$ of the bolometric output
\cite{1999ApJ...527..143Appleton,2000isdu.conf..232Appleton}.
The \mir\ emission is clearly enhanced in the nucleus and the expanding ring even though diffuse emission from the intervening medium was also observed
\cite{1999A&A...341...69Charmandaris,2001Ap&SS.276..553Charmandaris}. The diffuse components coincide with regions where only weak H${\alpha}$ had been previously measured. The \mir\ colours measured in collisional ring galaxies are similar
to those of late-type galaxies, indicative of more modest star
formation than is found for extranuclear starbursts (e.g. in the
nucleus inner ring and spokes of Cartwheel, the entirety of Arp10 and
the ring of Arp 118 \cite{2001Ap&SS.276..553Charmandaris}). The
morphological agreement between the \mir\ and H${\alpha}$ or radio emission is good although differences have been noted for VII Zw
466 \cite{1999ApJ...527..143Appleton}. A deviation from this picture is seen for the 
extranuclear starburst knot
in the archetypal collisional ring galaxy, the Cartwheel, that coincides
with the strongest H${\alpha}$ and radio knot. The regions has an unusually red \mir\
colour (15/7\mn$\sim$5.2) which is the highest among the extragalactic regions observed with \cam\ and it is even more extreme than the powerful starburst region (Knot A) revealed in the Antennae \cite{1999A&A...341...69Charmandaris}.
 Interestingly this region also hosts the
remarkable hyperluminous X-ray source (L$_{0.5-10keV} \gsim 10^{41}$ ergs/s \opencite{2003ApJ...596L.171Gao}) which dominates the X-ray emission from the
Cartwheel ring. 
\subsection{Hot star population, Starburst Evolution \& Metallicity}
\label{hotstars}
Starburst galaxies, defined as having star formation rates that cannot
be sustained over a Hubble time, are likely the birthplace of a large fraction of massive
stars. Locally, four starburst galaxies can account for 25\% of the
local massive star population within 10Mpc \cite{1998ApJ...503..646Heckman} and are important as the
providers of the Lyman continuum photons that ionise hydrogen.
\iso\ spectroscopy enabled the stellar populations of
starbursts to be probed using \fsl s that are sensitive to the nebular conditions from which they originate (see
\inlinecite{2001ApJ...552..544Foerster} for a detailed analysis of the stellar
populations and radiation environment of the local starburst
M\,82). The ratio of [NeIII]15.6\mn\ (Ep=41eV) and [NeII]12.8\mn\ (Ep=22eV) is sensitive to the hardness of the stellar energy distributions of the OB stars which excite them.
In a survey of 27 starburst galaxies,
\inlinecite{2000ApJ...539..641Thornley} confirmed earlier ground-based results that the excitation
measured by the [NeIII]15.6\mn/[NeII]12.8\mn\ ratio was lower in
starburst galaxies than measured for Galactic \hii\ regions, indicating a
deficiency of massive
stars in the starbursts. \citeauthor{2000ApJ...539..641Thornley} suggest a lower upper mass
cut-off to the initial mass function (IMF) or ageing of the massive
star population to be responsible for the low excitation. Through
extensive quantitative photo-ionisation modelling,
\citeauthor{2000ApJ...539..641Thornley} found the ratios to be
consistent with the {\em
formation} of massive stars (50-100M$_{\odot}$). This, together with the known presence of very massive stars in local starburst regions, led \citeauthor{2000ApJ...539..641Thornley} to prefer
an ageing scenario to explain the low excitation.
This explanation was further supported by modelling presented in
\inlinecite{2002ApJ...566..880Giveon}. The modelling of \citeauthor{2000ApJ...539..641Thornley} indicates a short starburst timescale of only a few O star lifetimes (10$^6$-10$^7$ yr) and suggests that strong negative feedback regulates the star-forming activity.
Disruption due
to stellar winds and SNe may cease the starbursts before the depletion
of gas and suggests that periodic starburst events are likely (e.g. in M82 \inlinecite{2001ApJ...552..544Foerster}). 

\inlinecite{2003A&A...403..829Verma} confirmed that for a
given nebular abundance, local starburst nuclei are of lower excitation than
Galactic and Magellanic cloud \hii\ regions. This was shown for the primary products of nucleosynthesis argon, sulphur and neon confirming the \inlinecite{2000ApJ...539..641Thornley} result was not
restricted to Ne only.
Additionally, the observed excitation was confirmed to be metallicity
dependent, with the lowest metallicity compact dwarfs showing the
highest excitations, at a similar level to local \hii\ regions.
Ageing, a modified IMF and metallicity are all possible contributors to
the metallicity-excitation relation. In the same study the abundance of sulphur was observed
to be lower than expected for a primary product of nucleosynthesis.
\wr\ and BCD galaxies did not show such an under-abundance
\cite{2003A&A...403..829Verma,2002ApJ...581.1002Nollenberg}.
This underabundance and weakness in the sulphur line strength implies
that (a) the \fsl s of neon are favoured over the those of sulphur as
a star formation tracer in future spectroscopic surveys of galaxies and
(b) that sulphur is an unsuitable tracer of metallicity.

\subsection{Wolf-Rayet Starbursts}
\label{wr}
The \wr\ phase is a signature of massive star formation as the progenitors are
believed to be massive young O stars ($3 \lsim t_{sb}
\lsim 8 Myr$, $M \gsim 20 M_{\odot}$). \wr\ features are found in a wide range of galaxies
such as BCD's, HII galaxies, AGN, LINERs and starbursts (see \opencite{1999A&AS..136...35Schaerer} for a compilation). Recently,
\inlinecite{2003MNRAS.340..289Lipari} identified possible \wr\ features in the spectra of some PG quasars. 
The identifying
\wr\ signatures are in the visible and thus are only representative of the
unobscured components. Nevertheless a correlation between the star
formation indicator UIB7.7\mn/15\mn\ continuum with H${\alpha}$ in \wr\ galaxies
suggests that the \ir\ and visible trace, at least partly, the same components (e.g. in Haro 3
\opencite{1996A&A...315L.105Metcalfe}). 
The short-lived
\wr\ phase is commonly detected in low metallicity systems with a strong
ISRF such as blue compact dwarfs (BCDs, see Sec. \ref{bcd}). 
\iso\ observations of known
\wr\ galaxies reveal differences in their \ir\ properties compared to
non-\wr\ galaxies. 
In comparison to starbursts, \wr\ galaxies are of higher excitation
and lower abundance contrary to expectation of a more important role
of \wr\ stars at higher metallicities from stellar evolution models
\cite{2003A&A...403..829Verma}. The \wr\ galaxy \n5253 
has a weaker excitation determined from \mir\ \iso\ data than expected by
stellar models \cite{1999MNRAS.304..654Crowther} implying that the UV
spectrum may not be as hard as commonly thought. However,
\inlinecite{1999MNRAS.304..654Crowther} show that the effect is due to
a young compact starburst surrounded by an older starburst both of
which lie within the SWS aperture. As a result, measured line
strengths are diluted and the apparent effective temperature of the
emitting region is lowered thus reducing the measured excitation.
This effect, that is fully consistent with an ageing/decaying
starburst scenario suggested by
\inlinecite{2000ApJ...539..641Thornley}, can explain the low
excitation of starbursts described in Section \ref{hotstars}
\cite{2000ApJ...539..641Thornley,2003A&A...403..829Verma}.

\subsection{Blue Compact Dwarfs}
\label{bcd}
A number of low metallicity (Z$<<$Z$_{\odot}$) blue compact ($<$5kpc) dwarf galaxies were
also observed by \iso.
These include I Zw 36
\cite{2001A&A...370..868Mochizuki,2002ApJ...581.1002Nollenberg}; Haro
11 \cite{2000A&A...359...41Bergvall}, Haro 3
\cite{1996A&A...315L.105Metcalfe}, \n5253
\cite{1999MNRAS.304..654Crowther}, II Zw 40
\cite{1999A&A...345L..17Schaerer,2000NewAR..44..249Madden}, \n1569
\cite{2000NewAR..44..249Madden,2002A&A...385L..23Haas,2003ApJ...588..199Lu,2003A&A...407..159Galliano},
\n1140 \cite{2000NewAR..44..249Madden}). 
Super-star clusters (SSCs) have been identified within some of them which are likely to be sites of massive star production that generate the bulk of the ionising flux \cite{2001ApJ...554L..29Gorjian,2002AJ....123..772Vacca,2004A&A...415..509Vanzi}). These clusters can also account for a substantial fraction of the emerging mid-infrared emission (e.g. \inlinecite{2002AJ....123..772Vacca}). The high obscuration, hot \ir\ emission and thermal radio spectra (measured for some) imply these are embedded young clusters. 
The properties of BCDs
are similar to low metallicity regions in the SMC
\cite{2000A&A...362..310Contursi}. 
Some BCDs also exhibit \wr\
features in their nuclear regions indicative of young massive stars
producing a significant amount of ionising UV photons (see Section
\ref{wr}).  \iso\ observations (see
\opencite{2000NewAR..44..249Madden} for a review) reveal varying characteristics between BCDs although \fsl s common
to starbursts are present e.g. [OIV], [SIII,IV] and [NeII,III] with ratios
that indicate high excitation (e.g. \opencite{1999MNRAS.304..654Crowther}).
The low metallicity and dust mass of BCDs in comparison to normal galaxies permits photoionisation on galaxy wide scales
\cite{2000NewAR..44..249Madden}. UIBs are weak or absent due to
destruction of their carriers by the intense radiation field and the
\mir\ is dominated by a strong VSG continuum.  [CII]/(CO) emission
(and [CII]/FIR to a lesser degree) can be greatly enhanced
($\times$10) in comparison to normal metallicity starburst galaxies
\cite{2000A&A...359...41Bergvall,2000NewAR..44..249Madden}. Along with the hot dust detected in BCDs
(e.g. \opencite{1996A&A...315L.105Metcalfe}), very cold dust
components were surprisingly observed in some BCD members of the Virgo
cluster
\cite{2002ApJ...567..221Popescu} and in NGC 1569 \cite{2003A&A...407..159Galliano}. In the latter, a very cold component (T=5-7K) contributes between 40-70\% of the total dust mass of the galaxy. Molecular gas (CO) and \ir\ emission indicate that, although relatively metal poor, these dwarfs have high dust and gas densities for their optical sizes \cite{2001A&A...370..868Mochizuki}. 

An exceptional BCD is SBS 0335-052, one of the most metal poor
galaxies currently known (1/41$Z_{\odot}$). This dwarf harbours
obscured, central SSCs
\cite{2001AJ....122.1736Dale,2001A&A...377...66Hunt} and shows a
young thermal radio spectrum
\cite{2003MNRAS.343..839Takeuchi,2004A&A...421..555Hirashita}. SBS
0335-252 is unusually bright in the \mir. No UIBs have been detected
and the [SIV] and [NeII] lines are weak in comparison to the strong
continuum
\cite{1999ApJ...516..783Thuan}. Estimates of the optical depth vary between authors from optically thin \cite{2001AJ....122.1736Dale} to A$_V\sim$10-30 \cite{1999ApJ...516..783Thuan,2002AJ....124.1995Plante}. The current optical, radio and \ir\ measurements make it difficult to present a consistent scenario for this intriguing galaxy. Further observations are needed to elucidate how such as small galaxy can produce such strong \ir\ emission without polluting its ISM with metals \cite{1999ApJ...516..783Thuan}. 
\subsection{Star Formation Rates}
\label{sfr}
The non-linear correlation (with large dispersion) between the \fir\ and
H$\alpha$ emission
in normal galaxies \cite{1992ApJ...396L..69Sauvage} suggests the use of
the \fir\ as a direct SFR estimator has limitations
\cite{1998ARA&A..36..189Kennicutt}.
At least for normal galaxies, both a contribution from the nuclear regions \cite{2001A&A...372..427Roussel} as well as from cold dust that
is unrelated to the starburst activity may
be responsible for the non-linear behaviour. 
At the highest \ir\ luminosities, such as in a sample of \uls, this
very cold component comprises only $\lsim$10\% of the total \fir\ flux \cite{2001A&A...379..823Klaas} . However, for less luminous sources that often have more
dominant very cold components (e.g the starbursts in
\opencite{2000ApJ...533..682Calzetti}), the contribution to the \fir\
from heating from non-ionising, older stars
should be considered when SFRs are
determined \cite{1987ApJ...314..513Lonsdale}.
A more basic problem
is that SFR estimates based on the \fir\ flux are prone to
errors arising simply from poorly determined \fir\ luminosities due to
inadequate sampling or constraints on the \fir\ SEDs. This affects often the faintest most luminous sources detected in \iras\ and \iso\ flux limited surveys where for some sources only one or two \fir\ data points are available.

The \mir\ also provides an alternative measure of the SFR. Based on \iso\ data for a
large sample of non-AGN-hosting galaxies,
\inlinecite{2001ApJ...549..215Dale} highlighted that, rather than the
commonly used \fir\ flux (42-122\mn), the \mir\ SED between 20-42\mn\
is most sensitive to the level of star formation activity.  Additionally,
the correlation of the \mir\ flux 7\mn/15\mn\
flux ratio with optical tracers of star formation such as H${\alpha}$ in
a range of galaxies (e.g. in normal galaxies
\cite{1996A&A...315L..89Sauvage,2001A&A...372..427Roussel}, Wolf-Rayet
galaxies
\cite{1996A&A...315L.105Metcalfe}, mixed pair local interacting
galaxies \cite{2003AJ....125..555Domingue,2005astro.ph..2459Domingue}, collisional ring galaxies
\cite{2001Ap&SS.276..553Charmandaris}, other interacting
galaxies \cite{2002A&A...391..417Lefloch}) and the \mir\ \hrl s, that
are correlated to the Lyman continuum, shows that the \mir\ flux can
be used as a SFR indicator
\cite{2001ApSSS.277..565Vigroux,2004A&A...419..501Foerster}. 
\inlinecite{2004A&A...419..501Foerster} provide an empirical
calibration of the UIB flux between 5-8.5\mn\ range as a SFR estimator which is valid
for starbursts up to the rate of M82, close to which the ISRF causes
sublimation of the UIB carriers. The VSG continuum represented by the
monochromatic 15\mn\ flux, that has been shown to correlate well with
the [ArII]6.99\mn\ line \cite{2003A&A...399..833Foerster}, can
withstand harder radiation fields providing a SFR estimator
for higher redshift, more luminous sources.
A potential scatter in this correlation can be introduced by an AGN
contribution to the VSG continuum; an event that may become more
common in galaxies of high luminosity (Sect. \ref{hls}).
In this case the \mir\ may not be a direct indicator of the SFR
unless the AGN contribution can be accurately subtracted
\cite{2000A&A...359..887Laurent}.

\subsection{FIR spectroscopy \& the [CII] deficit}
\label{cii}

The \fir\ spectrum of the starburst galaxy M\,82
\cite{1999ApJ...511..721Colbert} shows several \fsl s including those that are
commonly detected in active galaxies: [OI]63,145 \mn; [OIII]52,88\mn;
[NII]122\mn; [NIII]57\mn\ and [CII]158\mn. 
These features are
superimposed on a strong thermal continuum.
[CII]158\mn, along with [OI]63,145\mn\ and [NII]122\mn, are important cooling
lines of star forming regions and have been shown to correlate with
the dust continuum over 4 orders of magnitude in the Galaxy
\cite{2003eida.conf..209Baluteau}. For many galaxies, especially those
with low metallicities,
[CII]158\mn\ is the
strongest \fir\ line, corresponding to 0.1-1\% of the \fir\ flux
\cite{1997ApJ...491L..27Malhotra,2000NewAR..44..249Madden,2000A&A...359...41Bergvall}. 
The spatial 
correlation of the [CII]158\mn\ emission with obscured regions 
and the fact that it scales linearly with UIB and CO relative strength
suggests it originates in PDRs \cite{2000A&A...355..885Unger,1997ApJ...491L..27Malhotra,2001ApJ...561..766Malhotra,2001phso.conf..459Masegosa}. 
Fitting of the LWS spectrum of
M82 confirms this view with 75\% of the line 
produced in PDRs and the remainder from \hii\ regions
\cite{1999ApJ...511..721Colbert} while, for a sample of nearby galaxies, \inlinecite{2001A&A...375..566Negishi} find half to originate in PDRs and half in low density ionised gas.

The spectra of active galaxies range from those exhibiting strong
ionic fine structure emission lines (such as M82) to molecular
absorption line dominated systems (such as Arp 220)
\cite{1999Ap&SS.266...91Fischer,2000ibp..conf..239Fischer}.
The fine structure lines that are predominant in starburst galaxies are absent or weak in the spectra of higher luminosity \cite{1997ApJ...491L..27Malhotra,1997AAS...191.8903Fischer,1999Ap&SS.266...91Fischer,1999usis.conf..817Fischer}.
In particular, the [CII]158\mn/FIR continuum ratio decreases with increasing luminosity
\cite{1997ApJ...491L..27Malhotra} and in \uls\ is only 10\% of that measured for starbursts
\cite{1998ApJ...504L..11Luhman,2000ibp..conf..239Fischer,2003ApJ...594..758Luhman}
and is even lower in LINERs
\cite{2002AAS...20111615Sanei}. As an important cooling line of warm atomic gas, [CII] was expected to be high for strongly star-forming galaxies with a high UV radiation field. %
Contrary to intuition, the absence of the \fsl s and
the strong molecular absorption detected (e.g. OH, H$_2$O, CH, NH$_3$
and [OI]) imply a weaker radiation field in \uls\ in comparison to nearby
starbursts and AGN that is probably caused by shielding \cite{2000ibp..conf..239Fischer}. 

For sources deficient in [CII]158, the
[OI]63,145 \mn\ lines appear to become the preferred means of cooling star
forming gas
\cite{1999AAS...195.5311Malhotra,2000AAS...196.6305Brauher}.
In a systematic survey of LWS spectra of 36 local galaxies,
\inlinecite{2001A&A...375..566Negishi} found a strong decrease in the the [CII]158\mn/FIR and
[NII]122\mn/FIR cooling line ratios with bluer \fir\ colours. However no such trend was seen for [OI]63/FIR implying that the oxygen lines do not suffer from the [CII] deficiency.

A number of explanations have been proposed for the 
[CII] deficit.
Tracing [CII] towards the Sgr B2 complex,
which has an intense radiation field and gas density similar to active
galaxies, raises the possibility that self-absorption may contribute to the
observed deficiency
\cite{1999usis.conf..631Cox}. The [CII]158\mn\ line is seen both in
emission (originating from Sgr B2) and absorption
\cite{2002ApJ...581..315Vastel} consistent with the presence of intervening (cold)
clouds along the line of sight. This indicates the [CII]158\mn\ can be
optically thick. However, the weakness of the [OI]145\mn\ line observed
in Arp 220 and Mrk 231 and the large extinctions implied (A$_V\gsim$400, that exceed most other extinction measurements but see \inlinecite{2001A&A...367L...9Haas} for Arp 220) pose problems for solely a self-absorption explanation of
the [CII]/FIR deficiency 
\cite{2003ApJ...594..758Luhman}. Lower efficiency of photoelectric
gas heating for high radiation fields, such as those in the PDRs of
active galaxies, has been suggested
\cite{1997ApJ...491L..27Malhotra} but an accurate analysis of this scenario remains to be performed. 
Additionally, the radiation field and gas density may play a role. Ageing bursts 
were discounted 
because of the detection of the high excitation lines [NeIII] and [SIII] and because older starbursts are unlikely to fuel such high luminosities \cite{1999usis.conf..817Fischer}. Increased collisional
de-excitation, rather than the decrease in the photoelectric heating
efficiency, was supported by \inlinecite{2001A&A...375..566Negishi}.
\inlinecite{1998ApJ...504L..11Luhman} proposed high UV flux incident on
moderate density PDRs to explain the deficiency.
However, in a revision of this work, \cite{2003ApJ...594..758Luhman} preferred a non-PDR contribution to the \fir\ to explain the deficiency in the [CII]/FIR ratio. The authors suggest dust-bounded photoionisation regions as a possible origin since such regions contribute to the
\fir\ (i.e. increased UV-absorption by dust) but do not generate
significant amounts of [CII].
This
interpretation is corroborated by the increasing fraction of
dust-absorbed photons with ionisation parameter \cite{1992ApJ...399..495Voit}. This
scenario is also preferred in the radiative-transfer models of Arp 220,
although the effects of extinction cannot be ruled out
\cite{2004ApJ...613..247Gonzalez} .

The [CII] deficiency affects possible future searches of [CII]158\mn\
line emission from high redshift sources that have similar rest-frame
\fir\ colours to local \uls. Moreover, if the stronger cooling line in warmer \ir\ galaxies is [OI]63\mn, then the recent non-detections in four z=0.6-1.4
\ul-quasars \cite{2004ApJ...604..565Dale} also indicate a pessimistic view on either of these lines
as high-redshift tracers of active galaxies. The [OI]63\mn\ line may
also be self-absorbed.

\subsection{Gas-to-dust ratio and starburst evolution}
The gas-to-dust ratio is commonly used as an indicator of the
fraction of gas that is yet to be turned into stars. Dust mass
estimates determined from \iso\ SED analyses are used for this
determination. 
The gas-to-dust ratios
determined for starbursts, \uls, and quasars, are similar or
only slightly larger ($\sim$150-400) than those determined for the Milky Way ($\sim$150)
\cite{2000ApJ...533..682Calzetti,2003A&A...402...87Haas}.
The
highest gas-to-dust ratios are seen in BCD galaxies indicating young
stellar evolution \cite{2001AJ....122.1736Dale}. Low gas-to-dust
ratios and low dust temperatures indicate a post-starburst system
(e.g. the radio galaxy MG1019+0535
\cite{2001AJ....122..113Manning} and HyLIG F15307+3253 \cite{2002MNRAS.335..574Verma}). 

\section{Warm dust and hard radiation fields: nuclear AGN}
\label{hot}
\subsection{AGN in mergers}

\iso's low spatial scale could not discern emission from the torus and circumnuclear starburst regions even in the most nearby active galaxies.
As AGN emission is unresolved in \cam\ images, 
in this section we discuss the environment of the AGN, i.e. the heating of the ISM of the host
galaxy.
The collisional ring galaxies \n985 and Arp 118 both host Seyfert
nuclei. In \n985, \nir\ images resolve the bulge of a small, second
galaxy (whose accretion likely led to the formation of the ring) near
the active bright nucleus \cite{2002ApJ...566..682Appleton}.
The nuclei in both of
these galaxies show a hot \mir\ continuum that can wholly be attributed
to dust heating by active nuclei. UIB emission is suppressed
around the nuclei, suggesting they are absent in the vicinity of the
AGN. In the case of Arp 118, the Seyfert nucleus accounts for 75\% of
the \mir\ emission from the galaxy.
Two other \ls, VV 114 and Arp 299, show strong and hot \mir\ continuum
emission, i.e. that hot dust is present (a 230-300K blackbody describes the
VSG hot continuum), indicating that embedded AGN may
reside within these galaxies
\cite{2002A&A...391..417Lefloch,2004A&A...414..845Gallais}.
Seyfert
nuclei seem to have high \mir\ luminosities, as well as contributing
significantly to the total \mir\ emission of the system. %
In three merging pairs of \uls\ within which one member contains an AGN, the AGN hosting galaxy is invariably more \ir\
luminous than the other \cite{2002A&A...391..429Charmandaris}.
\inlinecite{1999Ap&SS.269..349Mirabel} note that the most luminous
sources are those that exhibit both AGN and starburst activity, reflecting the well known trend of increasing AGN importance with \ir\ luminosity
(see Sec. \ref{stagnulhy}). 

\subsection{Warm SEDs}
\label{warm}
\iso\ observations of the
12\mn\ \iras\ galaxy sample \cite{1993ApJS...89....1Rush}, that consists predominantly of Seyfert 1 and 2s
but includes starbursts and normal galaxies, displays a range of
SED shapes that can be expected for these source types (see Figure 2 in
\inlinecite{2002ApJ...572..105Spinoglio}). These
average SEDs show clear differences between normal galaxies, starbursts and
the Seyferts in the optical to \mir\ range, but they appear indistinguishable in the \fir.

By approximating the Seyfert SEDs using black-body functions, authors
such as
\inlinecite{1996A&A...315L.129Rodriguez} or \inlinecite{2002MNRAS.336..319Anton} found
that two components best fit their SEDs where the warmer component was
associated to the active nucleus and the colder component with star
forming or extended regions.
Similar associations were made for the two (or three) components implied by a novel inverse Bayesian method to analyse the SEDs of 10 Seyfert galaxies \cite{1998ApJ...500..685Perez}.
\inlinecite{2001A&A...377...60Prieto} showed that the fluxes of high excitation coronal lines detected in the \sws\ spectra of Seyfert galaxies, which are ascribed to excitation by the intense radiation field of the active nucleus, were found to be correlated to this \mir\ component but not to the \fir\ component. This implies that the \mir\ component arises from dust heated by processes related to the active nucleus. 

\subsection{Ionisation in AGN and Seyferts: the Big Blue Bump}
A thin accretion disk that fuels a black hole displays a
generic signature (quasi-thermal photons at T$\sim$10$^5$) in the
extreme UV (EUV) commonly called the big blue bump (BBB). The bump is likely to extend from the UV towards the soft X-ray bump but, due to galactic and intrinsic
absorption between the Lyman edge and a few hundred eV, the peak of
the bump is not directly detectable.
A novel method to reconstruct the BBB is to use the high ionisation potential \ir\ \fsl s (coronal lines, 50-300eV) that originate in the
narrow-line region of Seyferts and quasars and directly probe the EUV
SED. Reconstruction of the BBB was attempted using photo-ionisation modelling of the Seyferts \n4151(Sey 1.5), \n1068 (Sey 2) and
Circinus (Sey 2). The reconstruction is not only dependent upon the
ionising continuum but also on the geometry of the NLR and the
ionisation parameter. 
It was possible to
constrain the presence of this bump for these galaxies by
complementing the ISO data with UV, optical, and \nir\ lines. 
All three investigated sources
\cite{1996A&A...315L.109Moorwood,1999ApJ...512..204Alexander,2000ApJ...536..710Alexander}
require the presence of a thermal blue bump, a single power law
continuum fails to reproduce the observed narrow line ratios. Circinus
provides the clearest example of an EUV bump that peaks at 70eV and
contains half of the AGN luminosity. The predicted SEDs of \n4151 and
\n1068 display clear troughs falling sharply beyond the Lyman limit
and rising sharply above 100eV. The preferred explanation is that it is due to
absorption by neutral hydrogen of column density $N_{H}\sim 5 \times 10^{19} cm^{-2}$. In
the case of \n4151 this absorbing gas was identified through an
analysis of UV absorption lines \cite{1995ApJ...454L...7Kriss}.
This scenario allows for the presence of (obscured) BBB in all of
the sources investigated (see also
\inlinecite{2001A&A...377...60Prieto}).  However,
\inlinecite{1997A&A...327..909Binette} find that photoionization by a
simple power law suffices to explain Circinus, if an optically thin
(density-bounded) gas component is included. In a detailed metallicity
analysis, \inlinecite{1999A&A...342...87Oliva} find that either approach can equally explain
the observations, and note that the ionising continuum is mostly
dependent upon the gas density distribution. More complex geometry and
distribution of gas (e.g. 3-D multi-cloud models) and emission
mechanisms (shock heating and photoionisation)
\cite{1998ApJ...505..621Contini,2002A&A...386..399Contini,2003ApJ...587..562Martins}, than an assumed single cloud
with filling factor $<$1
\cite{1999ApJ...512..204Alexander}, are more appropriate assumptions
to model the narrow-line emission properties surrounding AGN
\cite{2003ApJ...587..562Martins}.

While such
detailed modelling was not possible for less well observed sources, a
comparison of spectroscopic tracers, such as electron density and
excitation, to photoionisation models enabled at least a consistency
check to be performed between power-law or black-body continua.
However, neither \inlinecite{2002A&A...393..821Sturm} nor \inlinecite{2001A&A...377...60Prieto} were able to distinguish the two
for their sources with the range of lines they
used. The metallicity, geometry and density of obscuring clouds and extinction all play a role in these analyses. 
\subsection{The dusty torus and variability} 
If near and mid-infrared emission from the
inner region of AGN is thermal re-radiation by dust surrounding the central
engine then, over sufficiently long timescales, variability seen in
the optical/UV should be detectable in the infrared. This would
provide convincing proof and constraints on the presence, temperature and distribution 
of dust grains from the variable source.
However
\iso\ monitoring of variable AGN, that were contemporaneous to other
multiwavelength campaigns, failed to show any convincing thermal
variability (e.g. Mrk 279 \opencite{2001A&A...369...57Santos}; BL Lac PKS 2155-304
\opencite{2000A&A...356....1Bertone};  BL Lac 2007+777 \opencite{2000A&A...353..937Peng}; OVV
quasar 3C 446 \opencite{2002PASA...19..152Leech}). Rather, the variability
where detected, was consistent with non-thermal components such as
jets \cite{2000A&A...353..937Peng,2002PASA...19..152Leech}.

These findings are consistent with variability traced by polarisation
measurements of the OVV quasar 3C\,279 at 170\mn\
\cite{1999ApJ...512..157Klaas} showing that the polarisation detected was aligned to the radio axis
and thus consistent with synchrotron emission.
The rapid variability
noted in two measurements separated by $\sim$1yr confirms that
the \fir\ emission is compact and originates from the core.
\subsection{Unification}
\label{unification}

Under the standard unification scheme
\cite{1993ARA&A..31..473Antonucci}, the orientation relative to the
line of sight of a dusty putative torus surrounding the accretion disk
of the central black hole, may connect various types of active
galaxies:
radio galaxies to quasars and Seyfert 1s to Seyfert 2s.
A check of these schemes is offered through their \ir\ emission: the dust torus
intercepts optical-UV photons released by the AGN and re-emits them in
the infrared. At $\lambda$$>$30\mn\ the \ir\ emission is largely optically
thin and hence isotropic. As a result the \ir\ SEDs should be similar for all
members of a unified class, irrespective of the orientation of the
obscuring torus to the line of sight. In this section we discuss how
\iso\ results have contributed to understanding this issue.
\subsubsection{Radio galaxies and quasars} \inlinecite{1982MNRAS.200.1067Orr} and \inlinecite{1989ApJ...336..606Barthel} proposed a scheme that unifies quasars and
radio galaxies as intrinsically similar sources, with observed
differences ascribed to the viewing-angle with respect to the black
hole.
A quasar with a flat radio spectrum is observed with a viewing- or aspect-angle of 0\deg\ to the line of sight (pole-on) such that a beamed jet with superluminal motion causes the bright
radio and \fir\ appearance (e.g.\ FR\,0234+28).  At intermediate angles between the pole- and edge-on orientations, a quasar with a steep radio
spectrum is seen. The shallower the aspect angle, the more obvious the
thermal bump of the torus becomes. At large orientations (90\deg\ or perpendicular) to the
line-of-sight, we observe the radio galaxy with the torus edge-on
(e.g.\ Cyg\,A, Fornax\,A, Centaurus\,A) whereby the nucleus that contains an ``optical
quasar'' is obscured by the dusty torus. Powerful Fanaroff-Riley
(FR)\,II radio galaxies are "edge-on" quasars viewed at high inclination
\cite{1998ApJ...503L.109Haas}. 
Overwhelmingly for samples of radio galaxies and radio quasars their
SEDs appear indistinguishable in the \fir\
\cite{1998ApJ...503L.109Haas,2000A&A...359..523Vanbemmel,2001A&A...372..719Meisenheimer,2002A&A...381..389Andreani,2004A&A...424..531Haas}
confirming the expectation from unified schemes i.e. that the dust is
isotropic and optically thin.
This is contrary to indications from
\iras\ data that implied radio galaxies have lower mid-far
infrared emission than radio quasars. The lower redshift range probed by \iras\ may account for this \cite{2001A&A...372..719Meisenheimer}. 
\subsubsection{Narrow line radio galaxies; broad line radio galaxies and
quasars} The 3C sample of quasars selected at 178MHz contains two
types of galaxies: (1) those exhibiting broad lines: steep spectrum
quasars and broad line radio galaxies; and (2) those exhibiting narrow
lines: FR\,II narrow line radio galaxies. Under the unification scenario, sources
of identical isotropic lobe power should have the same \fir\ isotropic
power. \inlinecite{2004A&A...424..531Haas} confirm earlier results of \inlinecite{2001A&A...372..719Meisenheimer} for {\em all}
3CR sources present within the \iso\ Data Archive\footnote{http://www.iso.vilspa.esa.es}, providing clear evidence in favour
of geometric unification. The considerable dispersion in lobe power
can be attributed to environment and
evolution. \inlinecite{2004A&A...421..129Siebenmorgen} confirm earlier work of \inlinecite{2003ApJ...599L..13Freudling} and find that the SEDs of
narrow-line radio galaxies to be colder ($\lambda_{peak}\sim100$\mn)
than the broad-line counterparts ($\lambda_{peak}\sim40$\mn). This
result is fully consistent with unification schemes where, in broad-line
radio galaxies, the BLR is exposed and so is hot dust that causes 
the SED shape to appear warmer.

\subsubsection{FR I and FR II galaxies} These extended radio sources are separated by the fact that FR\,II galaxies show powerful edge-brightened extended double lobes whereas FR\,Is have diffuse lobes that are brightest close to the nucleus. Moreover, FR\,Is are of lower luminosity than FR\,IIs. In a recent study of 3CR sources, \inlinecite{2004A&A...426L..29Muller} found that the low radio power FR I galaxies have low mid- and far-infrared luminosities.
In contrast, the FR II sources are \ir\ luminous. The high \fir-to-\mir\ ratio 
and the temperature of the \fir\ peak which is comparable to that of quiescent radio-quiet elliptical galaxies, implies that heating by the ISRF of the host galaxy is probable. This together with the finding that the dust masses of FR\,Is and FR\,IIs are similar but with low gas fractions in the nuclear regions, implies that the black holes in FR\,I galaxies are fed at a lower rate than FR\,II galaxies \cite{2004A&A...426L..29Muller}.

\subsubsection{Flat spectrum and Steep spectrum Quasars} Flat spectrum quasars
show a strong synchrotron component (e.g. FR 0234+28), smoothly
concatenating the mm flux to that observed in the \mir\
\cite{1998ApJ...503L.109Haas}. A thermal
\fir\ component may equal the synchrotron in strength (e.g. 3C\,279),
but in most cases only upper limits to a thermal contribution can be
derived due to the dominating synchrotron emission. This prevalence of
\fir\ synchrotron emission is consistent with Doppler boosted
synchrotron emission of a relativistic jet seen almost end-on and does
not contradict this unification \cite{1998ApJ...503L.109Haas}.

\subsubsection{Radio-loud and radio-quiet quasars}
A complete range of radio-loud, radio-intermediate and radio-quiet
quasars was analysed by
\inlinecite{2000A&A...362...75Polletta}. Except for flat radio sources,
the mid- and far-\ir\ emission is of thermal origin.
The \fir\ emission is predominantly explained by heating from the AGN itself and, while starburst components are present, they contribute to the \ir\ emission at different levels but always less than the AGN
($\lsim$27\%).
\subsubsection{Seyfert 1s and Seyfert 2s} Within the unified scheme, Seyfert 1s
and 2s differ only in viewing angle of the torus obscuring the central
black hole. In Seyfert 1s the broad line region is exposed whereas in
Seyfert 2s it is hidden from view due to obscuration by the torus. The
central regions where dust is heated to high temperatures gives rise
to a \mir\ peak in the continuum emission. Aspect-oriented unification
is consistent with the result that the \mir\ dust temperatures of
Seyfert 1s are higher than in Seyfert 2s, where the torus is partially
optically thick up to the
\mir\ \cite{1998Ap&SS.263..103Perez}. Spectroscopic tracers such
as high excitation lines and broad lines are indicative of the presence
and orientation of the putative torus. In a large sample of Seyferts,
\inlinecite{2000A&A...357..839Clavel} showed that UIB equivalent widths are larger in Seyfert 2s than in Seyfert
1s. An interpretation of this result is that UIB
features originate from extended regions and hence emit isotropically,
while the AGN continuum is absorbed by the
torus and must emit anisotropically \cite{2001ApJ...557...39Perez}
since it is dependent upon the orientation of the torus to the line of
sight.
This interpretation thus provides direct
proof of the unification of the Seyfert classes. 

A test of unification schemes is offered via observations in hard
X-rays that can penetrate an obscuring (Compton-thin) torus. Thus, in
the simplest unification models, if X-ray emission is not beamed,
the \mir\ continuum arising directly from AGN heating of
the torus or dust in the narrow line region is determined by the geometric
distribution of the obscuring material and the aspect angle.
After isolating the
\mir\ continuum due to the active nucleus alone,
\inlinecite{2004A&A...418..465Lutz} found no distinction between Sey
1s and 2s with the X-ray/\mir\ flux ratio, contrary to the predictions
of unified schemes. The authors suggest the apparent uniformity is due
to the dilution of anisotropic emission by a predominant \mir\
component on extended scales that emits isotropically.  

Another result that is difficult to reconcile with the torus model
arises from \iso\ SED analysis of a sample of hard X-ray selected
sources comprising both type 1 and 2 Seyfert
galaxies. \inlinecite{2003ApJ...590..128Kuraszkiewicz} showed that
while Seyfert 2s are redder in their optical colours than Seyfert 1s,
their mid- to
\fir\ emission, as traced by the 25 to 60\mn\ colour, remains similar.
More complex
geometries for the obscuring material are required to explain both the
difference in optical colours and the similarity in the \ir.  Clumpy and
extended models are suggested where a disk of only a few hundred
parsecs can reproduce the IR SED without invoking a starburst
component (see \opencite{2003astro.ph..9040Elitzur} and Sect. \ref{stagn}).

\begin{figure}
\centerline{\includegraphics[width=25pc]{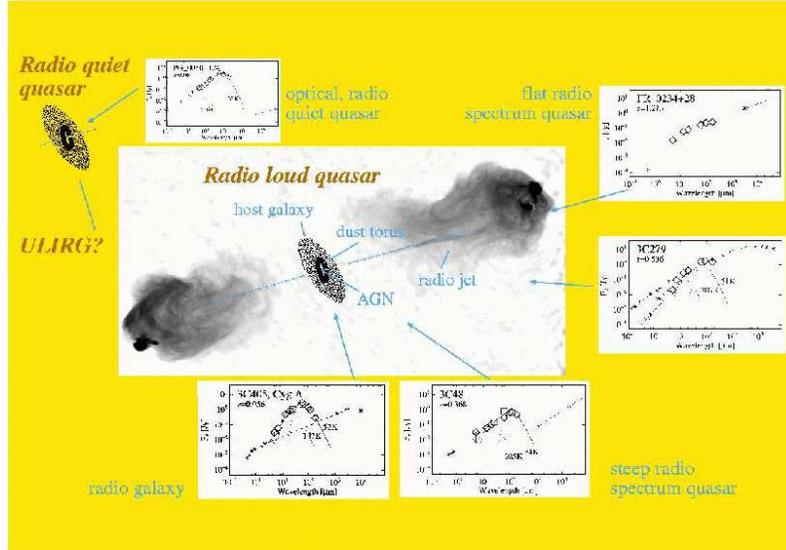}}
%\centerline{\includegraphics[width=25pc]{VermaA_3_small.pdf}}
%
\caption{Illustration of the unified scheme of quasars complemented by its manifestation in the \ir\ SEDs of quasars (Haas et al. 1998).}
\label{unifiedhaas}
\end{figure}

\subsection{Evolution of compact steep spectrum radio galaxies}
\iso\ data contributed towards understanding whether the compactness of
steep-spectrum radio galaxies was due to age or evolution of
environmental conditions that inhibits the growth of the radio
lobes. It was expected that dense material causing this inhibition should be
accompanied by \fir\ emitting
dust. However, \inlinecite{2000A&A...358..499Fanti} found no \fir\ excess in
compact steep spectrum radio galaxies, in comparison to a control sample of 16 extended
radio galaxies, which favours a youthful scenario to explain
their compactness.

\subsection{Constraints from line profiles}

\subsubsection{Probing the NLR}
The emission line profiles of Seyfert galaxies are known to exhibit blue asymmetries and blue-shifts with respect to the systemic velocity in their optical spectra. The profiles provide constraints on the NLR.
Narrow lines
detected in the \mir\ have the advantage of being much less sensitive to extinction than optical lines and therefore provide a relatively unobscured view to the
NLR. The similarity in line shapes between the optical and the \mir\
provides constraints on the nature of the NLR and models how such
asymmetries were produced. \inlinecite{1999ApJ...512..197Sturm} are able to rule out simple radial flows plus dust scenarios and they propose that a central
geometrically thin, but optically thick, obscuring screen located close
to the nucleus is present in \n4151 to explain the observed asymmetry. The \mir\ line profiles of \n1068 are due to a combination of an highly ionised outflow and extended emission of lower excitation, and remaining blue shifts attributed to an intrinsic asymmetry in the NLR or very high column density in the line of sight \cite{2000ApJ...536..697Lutz}. 
For a larger sample, \inlinecite{2002A&A...393..821Sturm} show that the differences in line profiles depend upon the 
ionisation potential.
Circinus shows symmetric low-ionisation and
asymmetric high-ionisation lines while the spectrum of NGC7582 shows
the reverse.
These differences suggest that the NLR properties affecting
individual galaxies are particular to that galaxy. Differences
including extinction, asymmetric distributions of NLR clouds,
orientation must all influence the line shapes
\cite{2002A&A...393..821Sturm}.  

\subsubsection{Probing the BLR}
\inlinecite{2000ApJ...530..733Lutz} searched for broad components in the
\hrl s of \n1068, since the detection would imply that at least part of the
BLR is accessible in the \mir\ (looking through the obscuring putative
torus). The lack of detection places a lower limit in both extinction ($A_{V}>50$mag) and hydrogen column density
($N_{H}>10^{23}cm^{-2}$). Although little can be stated regarding
favoured torus models or constraining gas properties based upon this
information, Br$\alpha$ was identified as being the preferred
indication for future studies and \inlinecite{2002A&A...396..439Lutz} found broad line
components in approximately 1/4 of their sample of Seyfert galaxies from
ground based high resolution spectroscopy.

\section{Ambiguous Sources: Ultra- \& hyperluminous \ir\ galaxies}
\label{stagnulhy}

As we have previously discussed, \ul\ samples contain galaxies of both starburst and AGN type. Establishing whether these types are linked in evolution or nature is a key issue in the study of active
galaxies raising several questions: what is the role of star formation and black holes; which is/are
present; do super-massive black holes and starburst activity concurrently exist in all \uls\ and how much do they contribute
to the bolometric luminosity?
Prior to
\iso, few observational tools were available to probe through the dust obscuration to reach the source of the \ir\ power.
\iso's spectroscopic and photometric capabilities enabled significant
advances in this field, largely elucidating the previously
termed 'starburst-AGN controversy' which is discussed in this section.

\subsection{Quantitative Spectroscopy}
\label{diagnostics}
\subsubsection{Fine Structure Line Diagnostics}
\begin{figure}
\tabcapfont
\centerline{
\begin{tabular}{c}
\includegraphics[width=18pc]{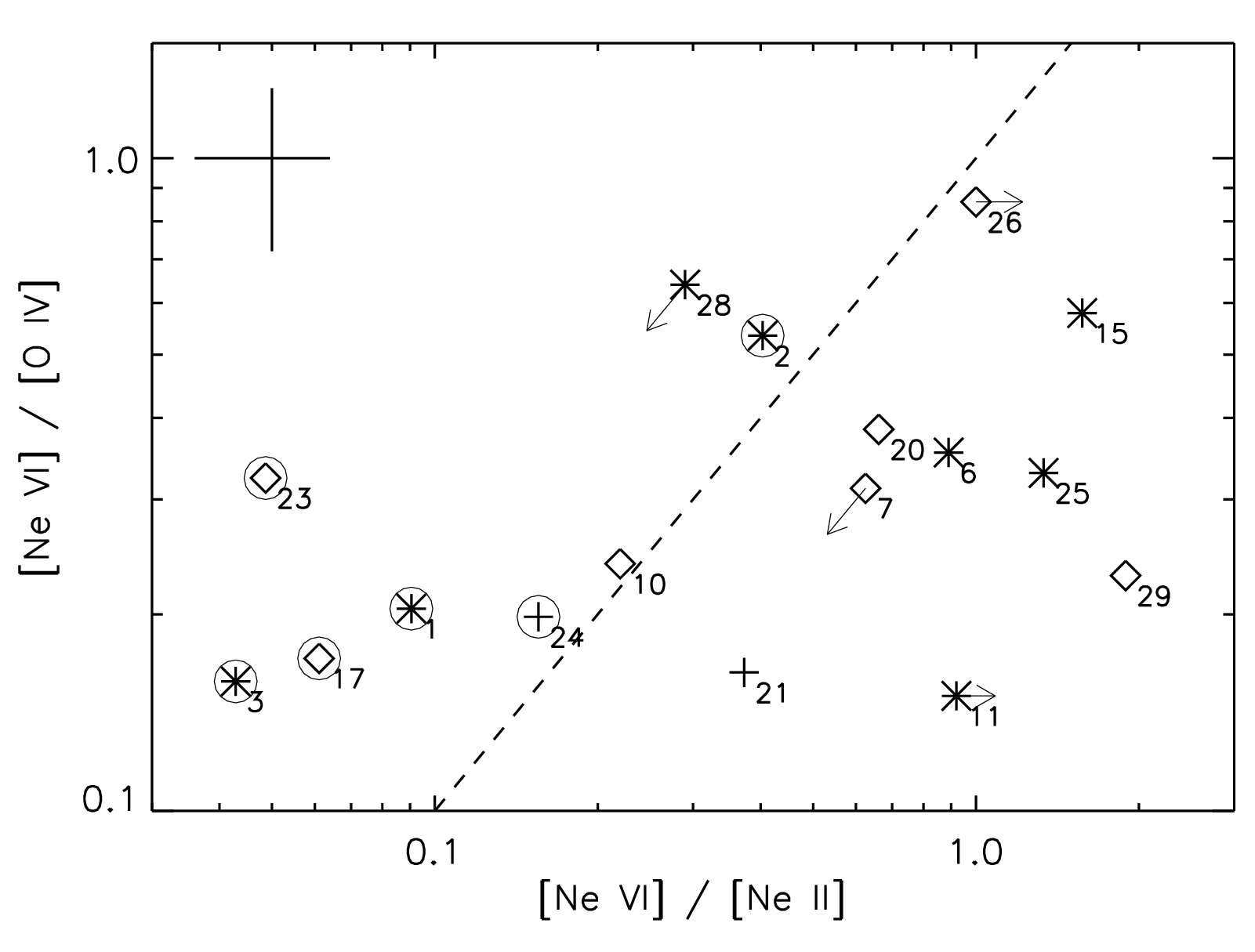}\\ 
fig. a\\
\includegraphics[width=18pc]{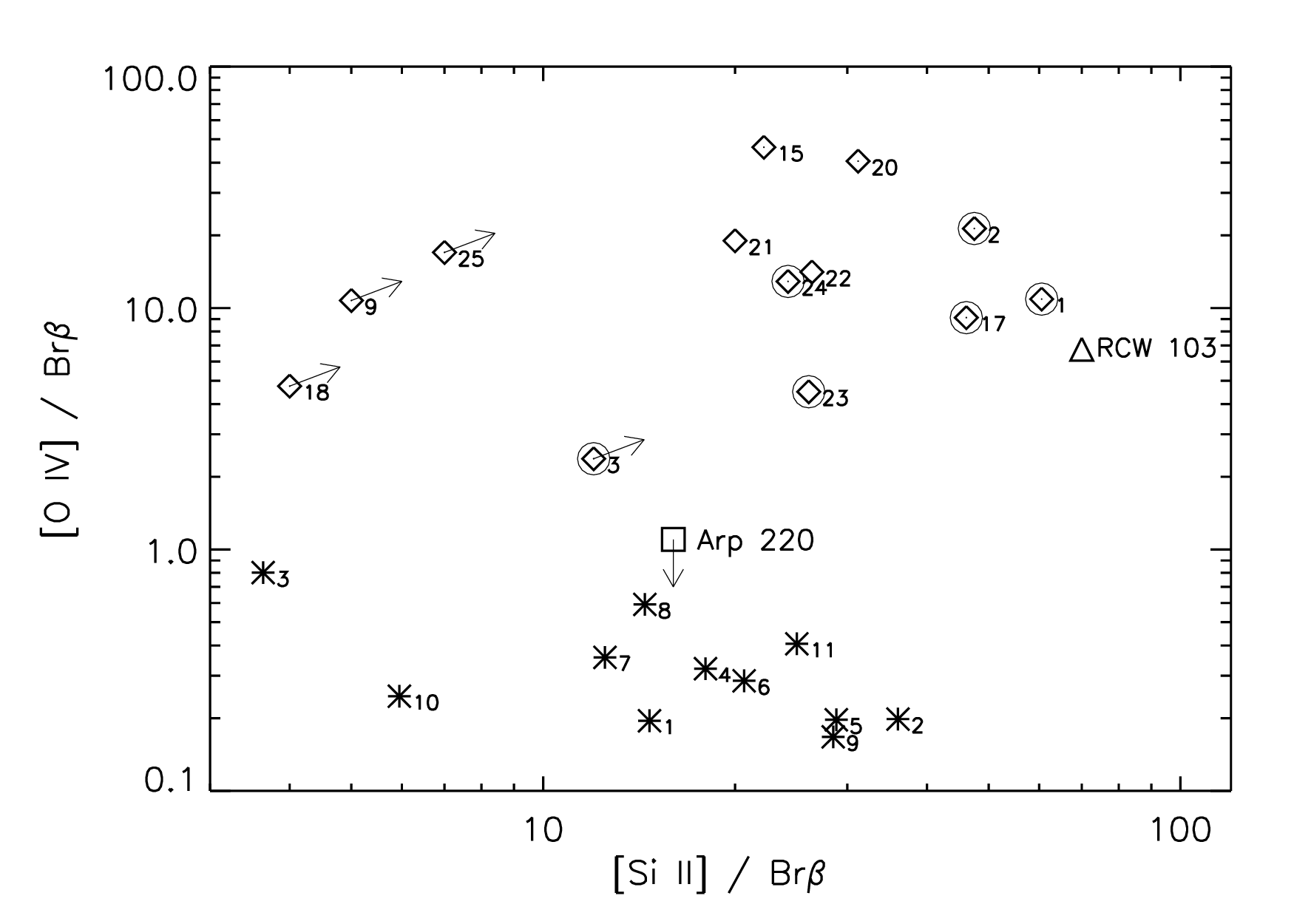} \\
fig. b\\
\end{tabular}}
\caption{Diagnostic diagrams based upon \fsl s from Sturm et al. (2002). Fig. a: [NeVI]/[OIV] vs [NeVI]/[NeII]: Diamonds - Sey 1s, Asterisks - Sey 2s:  the composite Seyfert galaxies that exhibit UIBs are encircled and are mostly separated from the remaining galaxies. Fig. b: [OIV]26\mn/Br$\beta$ vs [Si II]34\mn/Br$\beta$: Diamonds - AGN asterisks - Starbursts: the starbursts occupy a different regions of the diagnostic diagram than the AGN.%
}
\label{sturm}
\end{figure}

Fine structure lines directly trace the excitation states of the ISM from which they originate. The radiation fields generated by AGN/QSO greatly exceed that of starbursts and thus can excite interstellar gas to higher ionisation species (as described in Sect. \ref{fsl}). 
The difference in excitation between AGN and starbursts is the
basis of \fsl\ diagnostic diagrams. Such excitation
diagrams are well established in the optical
\cite{1987ApJS...63..295Veilleux,2001ApJ...556..121Kewley}, for which
\inlinecite{2002A&A...393..821Sturm} constructed the first infrared analogues (Fig. \ref{sturm}) that successfully distinguish starbursts from AGN.

Diagnostic diagrams from \inlinecite{2002A&A...393..821Sturm} are shown in Fig. \ref{sturm} both involving the \fsl\ [OIV]25.91\mn.
While this line was expected in the high ionisation
potential of AGN, it was also often weakly detected in star-forming
galaxies and ULIGs \cite{1998A&A...333L..75Lutz}. In our Galaxy, the [OIV]25.91\mn\ line has not been
observed in \hii\ regions surrounding young, hot, massive stars. However,
its detection in \iso\ spectra of starbursts and the fact that it was
spatially resolved in the low excitation starburst M82 implies it is
plausibly produced by ionising shocks or extremely hot ionising stars
\cite{1998A&A...333L..75Lutz,1999A&A...345L..17Schaerer}. 
Its
diagnostic capability thus arises from the fact that the measured
strength of the line in starburst galaxy spectra is at least ten
times fainter than those measured in spectra of AGN.  This absolute
difference in the strength of [OIV] relative to a low excitation starburst line such as
[NeII] provides a straightforward indicator of AGN versus starburst
activity \cite{2002A&A...393..821Sturm}.

Combining the high:high excitation \fsl\ ratio [NeVI]/[OIV] with the
high:low excitation \fsl\ ratio [NeVI]/[NeII], provides a basic
diagnostic of composite sources (Fig. \ref{sturm}a).
By using the limiting
values of the high:low excitation ratio [OIV]/[NeII] (always $<$0.01
in starbursts and 0.1$<$AGN$<$1.0) %
\inlinecite{2002A&A...393..821Sturm} were able to estimate the fractional contribution of an AGN to the bolometric luminosity through a simple mixing model. Fig. \ref{sturm}b shows how the high and low ionisation indicators normalised to the \hrl\ Br$\beta$ (an indicator of the Lyman continuum rate) provides an analogue to the \inlinecite{1987ApJS...63..295Veilleux} diagnostics that are insensitive to obscuration and will be extremely useful for classifying \uls\ observed with future \ir\ spectrographs. An equivalent diagnostic at longer wavelengths can be constructed using 
[CII]158\mn/[OI]63\mn\ versus [OIII]88\mn/[OI]63\mn\ 
that can separate starburst, AGN and PDR contributions
\cite{2003agnc.conf..557Spinoglio}.

As the spectroscopic sensitivity of \iso\ permitted the detection
of \hrl s or useful limits on the [OIV] emission in only the brightest
sources, alternative tools using more commonly detected spectral features (e.g. UIBs and bright \fsl s) were used during earlier \iso\ science analysis.
\inlinecite{1998ApJ...498..579Genzel} combined excitation and UIB strength to asses the nature of a large sample of AGN, starbursts and \uls. \inlinecite{2000A&A...359..887Laurent} constructed a diagnostic based upon the \mir\ continuum shape to UIB strength (Sect. \ref{uibdiag}).
However the use of UIBs is limited and presents problems in
interpretation (Sect. \ref{uiblimit}).

\subsubsection{UIB-based diagnostics}
\label{uibdiag}
The predominance of UIBs in starbursts (and their deficiency in AGN,
or low metallicity environments) naturally lends the ratio of UIB
feature flux (normalised to the \mir\ continuum) to be a suitable
tracer of star formation activity
\cite{1996A&A...315L.337Verstraete,2004A&A...419..501Foerster,2004ApJ...613..986Peeters}.
For normal and starburst galaxies the UIB to continuum ratio is seen
to be high in the nuclear star-forming regions and to decrease with
radial distance
\cite{1999A&A...342..643Mattila,2000AJ....120..583Dale}, indicating a
weakening ISRF. Conversely, a strong enhancement in the continuum to
line ratio is seen in the vicinity of AGN, where the continuum is
strong and UIBs weak as the features suffer dilution due to the
absolute high level of the continuum and/or destruction of their
carriers \cite{1998ApJ...505L.103Lutz}.
The effectiveness of this indicator is clearly seen in Centaurus A,
where spatially resolved \cam-CVF observations of the AGN-dominated nucleus and the
extended starburst region \cite{1999A&A...341..667Mirabel} show
striking differences in their \mir\ emission. Together with the
absence of UIB emission features, the spectral index of the continuum
is steeper in AGN than in the surrounding circumnuclear star forming
regions.  These spectral differences enable, at least in principle,
discrimination between the emission characteristics of the two
distinct physical mechanisms.

\begin{figure}
\tabcapfont
\centerline{
\begin{tabular}{c}
\includegraphics[width=20pc]{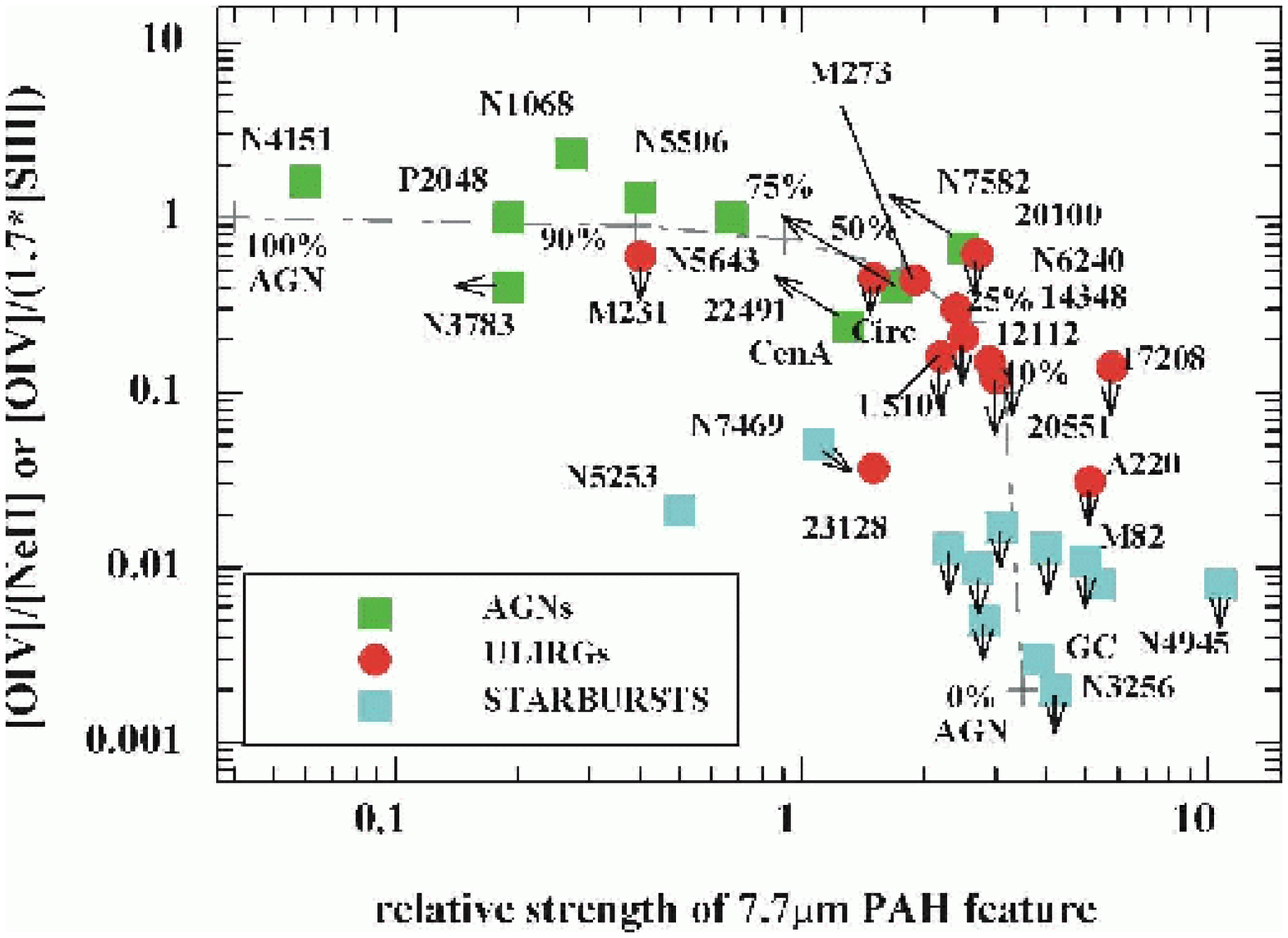}\\
Fig a\\
\includegraphics[width=20pc]{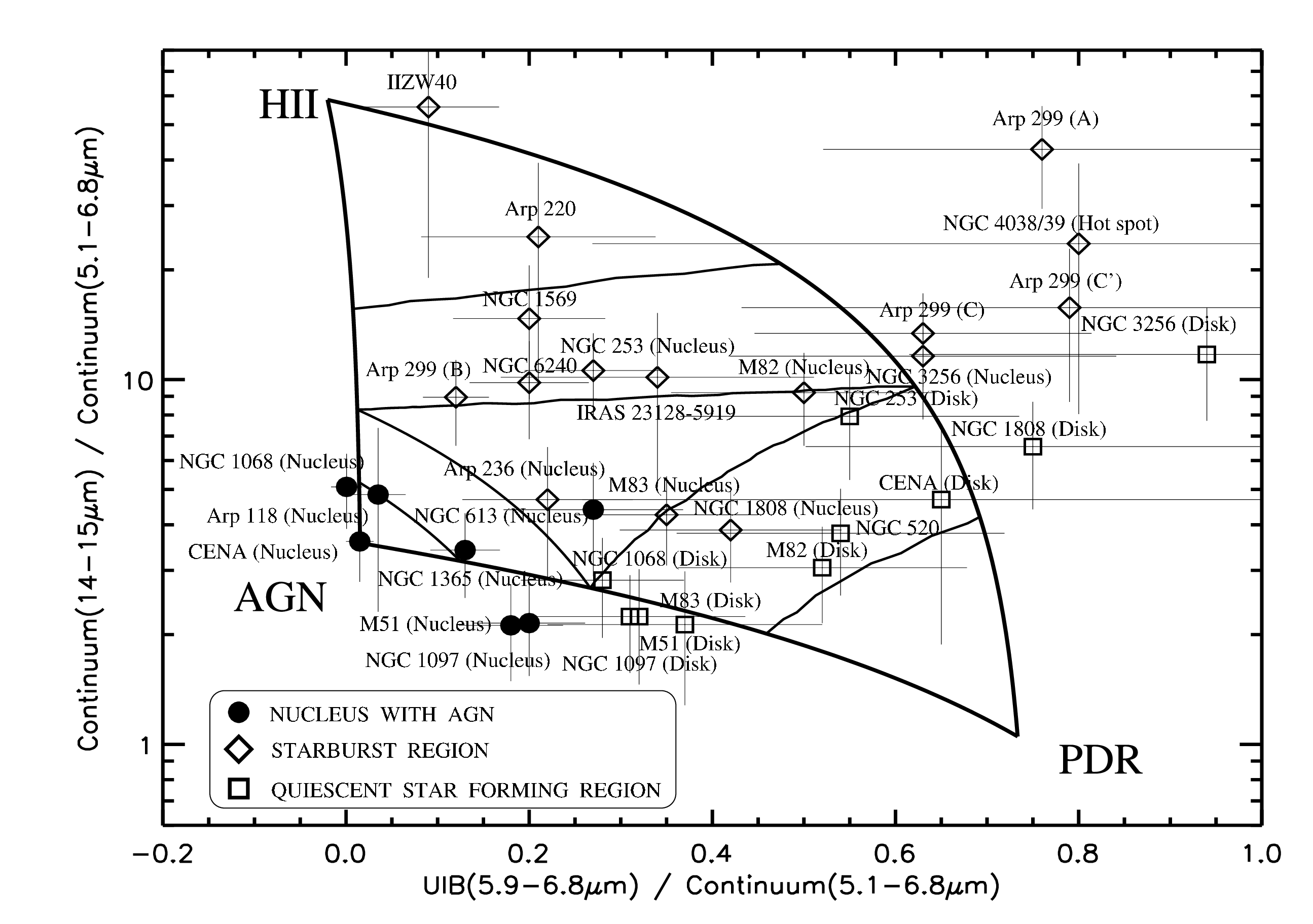}\\
Fig b\\
\includegraphics[width=20pc]{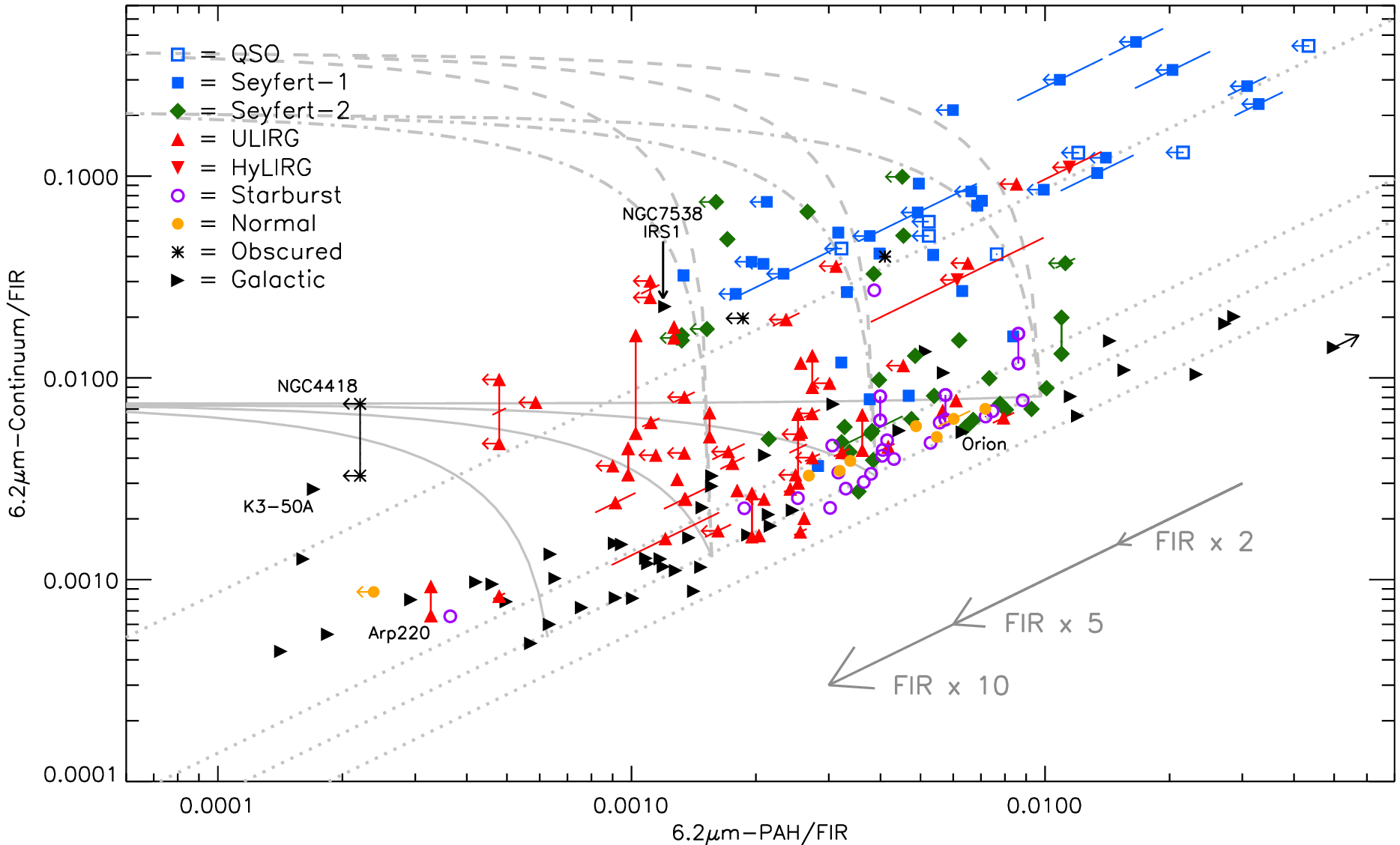}\\
Fig c\\
\end{tabular}}
\caption{UIB diagnostics from Genzel et al. (1998) (a) and Laurent et al. (2000) (b). Both of these diagrams show the mixing levels of starbursts and AGN in active sources. Finally the MIR/FIR diagnostic of Peeters et al. 2004 is shown for comparison (c).
}
\label{genzel}
\end{figure}

\inlinecite{1998ApJ...498..579Genzel} demonstrated the effectiveness of the
7.7\mn\ UIB-to-continuum ratio versus a tracer of hardness (such as
the ratio of the high excitation potential \fsl\ [OIV] to the low
excitation line [NeII] (or [SIII])) to classify
ambiguous sources. This 'Genzel-diagram' (Fig. \ref{genzel}a)
enabled the classification of \uls\ for which the underlying fuelling
mechanisms were unclear in the pre-ISO era. The separation between
AGN, starbursts and the intervening \uls\ is unequivocally depicted.
The well known low-excitation starburst M82 is far separated from the
AGN NGC 4151. These extremities more or less define an appropriate
mixing line that begins with 0\% AGN in the lower right corner to
100\% in the upper right. Thus placing a \ul\ on this plane provides an
indication of its AGN fraction. The line of equivalent mixing is
marked on the diagram, 4 of the 13 \uls\ shown fall to the left of
that line implying a bolometrically dominant AGN is needed to explain
their emission.

Similarly, using the ratios of 6.2\mn\ UIB-to-continuum vs. warm-to-hot
continuum, \inlinecite{2000A&A...359..887Laurent}
constructed a diagnostic diagram that can quantitatively estimate the
fractional contribution from AGN, \hii\ regions and PDRs.
\inlinecite{2000A&A...359..887Laurent} showed how a \mir\ spectrum may
be quantitatively decomposed into these three components through
spectral template fitting (see Fig. \ref{genzel}b), using the pure \hii\ spectrum of M17 to fit the strong VSG continuum in starbursts \cite{1996A&A...315L.305Cesarsky}, an isolated PDR spectrum of the reflection nebula \n7023 \cite{1996A&A...315L.309Cesarsky} and the nuclear spectrum of Centaurus A representing the hot AGN spectrum \cite{2000A&A...359..887Laurent}. Such diagnostics are extremely valuable for
assessing the fuelling mechanisms of ambiguous sources such as \uls\
\cite{2000A&A...359..887Laurent,2001ApJ...552..527Tran} and can reveal
the presence of obscured AGN in the \mir. This diagnostic tool is most
successful when used with spectra of high signal-to-noise and with sufficient coverage of the \mir\ range [i.e. to best determine the continua shortward ($<$5\mn) and longward ($>$12\mn) of the family of strong UIB features] 
that enables accurate
decomposition of the \hii\, PDR and AGN templates and reduces
problems associated with blended \mir\ absorption and emission lines
\cite{2002A&A...385.1022Spoon}. A follow up of this method, on a
larger sample of galaxies was presented by
\inlinecite{2004ApJ...613..986Peeters} with similar results. The authors also construct a new mid- and far-\ir\ combined diagnostic that is useful for differentiating obscured sources (Fig. \ref{genzel}c). In integrated galaxy spectra, where no
spatial information is available, e.g. for compact and/or distant
sources, decomposition of the spectral characteristics is the
only means to identify the power sources constituting the \mir\
emission (e.g. \opencite{1997eai..proc...63Vigroux},
\opencite{1999AJ....118.2625Rigopoulou};
\opencite{1998ApJ...505L.103Lutz}; \opencite{1998A&A...334..482Mouri};
\opencite{1998ApJ...498..579Genzel};
\opencite{2000A&A...359..887Laurent}).

\subsubsection{Limitations}
\label{uiblimit}
The use of the UIB strength as a diagnostic has several limitations.
One originates from the fact that
measuring the strength of the 7.7\mn\ UIB feature in the wavelength
range of \pht-S (2-11\mn) was challenging as ULIGs also display
strong silicate absorption at 9.7um and tracing the continuum was
often extinction dependent. 
Another limitation of relying on UIB strength originates from its
sensitivity to the radiation environment. 
Its absence in hard radiation fields in the vicinity of massive stars 
or AGN weakens its effectiveness as a tracer of star formation. 
Finally, as discussed by Laurent et al.
(\citeyear{1999usis.conf..913Laurent,2000A&A...359..887Laurent}), \iso's limited
spatial resolution in the \mir, may introduce a scatter into the
diagnostic diagram, especially for distant sources with angular size
smaller than the instrument beams. This results in a dilution of the
AGN signatures as a larger fraction of the circumnuclear star forming
regions of the host galaxy disk enter into the beam.
For these reasons, while UIB diagnostics were relatively
successful in particular for faint sources. The diagnostics formed purely from \fsl s, which directly trace differing
excitation regions within galaxies, offer far cleaner and more
effective means of source-type segregation and should be preferred for classification of faint and high-redshift galaxies that will be observed by future telescopes.
\subsubsection{Broad-band colour diagnostics}
The well known diagnostics based on the broad
band infrared colours determined from \iras\ data (e.g. the \iras\
60\mn/25\mn\ colour represented the warm/cold emission ratio) have
been extended to include \iso\ photometry
\cite{2002ApJ...572..105Spinoglio} 
and show the separation of 'warm' AGN (Sect. \ref{warm}), 'cold' starburst (Sect. \ref{cold}) and 'very cold' (Sect. \ref{verycold})
dust dominated systems (see their Fig. 14). While in the \fir, the SEDs
of Sey 1s and 2s are indistinguishable, the diagram of
\inlinecite{2002ApJ...572..105Spinoglio} is based on the fact that the
SEDs of Sey1s are flatter in the mid- to far-\ir\ than Sey 2s.
Using averaged spectra from this sample, the authors were able to
produce redshift tracks on colour-colour diagrams for the Spitzer MIPS
and IRAC filters as a means of determining \ir\ photometric
redshifts. Even though these tracks were calculated without
evolutionary considerations, the diagrams are useful tools to
differentiate AGN (Seyfert 1 \&2s, QSOs) dominated sources from
starbursts.
\subsection{The nature of \uls}

Using these techniques, the majority of \uls\ are found to be
predominantly powered by starbursts. For large samples of \uls\
\inlinecite{1998ApJ...505L.103Lutz},
\inlinecite{1999AJ....118.2625Rigopoulou} and
\inlinecite{2001ApJ...552..527Tran} corroborate the early result of
\cite{1998ApJ...498..579Genzel} and find that $\sim$75-85\% of their
sample do not contain energetically dominant AGN. Starburst activity
can explain the bulk of the bolometric luminosity although the
presence of a weak or heavily obscured AGN cannot be excluded. While
the fraction of AGN dominated sources is 15-25\%, the fraction of AGN
{\em hosting} \uls\ is 20-25\%. The AGN dominated sources have 'warm'
\iras\ colours ($S_{25}/S_{60}\gsim0.2$) confirming previous \iras\
results 
on \uls\ with optical or \nir\ spectroscopic classification.
Moreover, a comparison of the
\mir\ classification with good quality optical spectroscopy is
remarkably consistent 
providing both optically classified \hii\
regions and LINERs are both ascribed to the \mir\ starburst
class
\cite{1999ApJ...517L..13Lutz}.  
\subsection{The nature of LINERs}

The nuclei of a large fraction (as much as one-third) of local
galaxies display spectra with weak emission lines of lower ionisation
than is seen in typical AGN, e.g. Seyfert nuclei. Furthermore, the low
ionisation species present can not be explained through photoionisation
by normal stars.
The population of low ionisation nuclear emission-line regions - LINERs
\cite{1980A&A....87..152Heckman} is well known to be mixed. 
Because of their prevalence, establishing the fraction of AGN or
star-forming like LINERs has strong implications on our understanding of galaxies.
\iso\ spectroscopic results clearly indicate that, at
least in the infrared, their properties are more similar to starbursts
\cite{1999ApJ...517L..13Lutz}. In this case, the low ionisation emission lines are ascribed to ionising shocks related to starburst driven winds \cite{1999ApJ...517L..13Lutz,2000ApJ...529..219Sugai}. Interestingly though, in an \ir\ study of LINERs, \inlinecite{2004A&A...414..825Satyapal} found as much as two-thirds of the sample
with hard X-ray counterparts
have compact X-ray morphologies
consistent with those found for AGN. The remainder with
scattered X-ray emission are brighter in the \ir\ than the
compact ones. The authors found that most LINERs displaying 
high excitation
\mir\ lines also show compact X-ray morphology, even though a few systems deviate from this 
(i.e. \n404 and \n6240). These studies indicate that \ir\ bright
LINERs are most likely members of a shock-dominated population whereas
optical selection may trace the low luminosity AGN component. The fact
that LINERs occupy a locus in ionisation diagrams situated between
classical starbursts and AGN suggests
an evolutionary connection whereby LINERs may represent a phase when
the AGN is deeply buried and the unrelated starburst dominates the
detected emission \cite{1999ApJ...517L..13Lutz}.
\subsection{ULIG SEDs}
An alternative classification method arises from an analysis of \ul\ SEDs
that include \iso\ data. This 
includes fainter and/or more luminous \uls\ and \hls\ that were
inaccessible to \iso 's spectrometers.  Often in conjunction with
multi-wavelength data from other telescopes, \iso\ data has enabled
classification of \uls\ through decomposition of their SEDs.
A statistical study of 20 \uls\ \cite{2001A&A...379..823Klaas} clearly
showed a progression of SED shapes. These were broadly divided into
two types. The first (type A) show a power law increase from 1 to
50\mn\ and are classified as Seyferts from optical spectroscopy. They
also have warm \ir\ colours (f$_{\rm 25 \mu m}$\,/\,f$_{\rm 60 \mu m}$
$>$ 0.2). Type B sources display a relatively flat SED from 1 to 10\mn,
that is followed by a stepp rise towards the peak of emission in the
\fir. These SED characteristics are displayed by \uls\ with optical
spectroscopic classifications of Seyfert 2, LINER,
\hii/starburst. These are 'cool' \uls\ with \ir\ colour f$_{\rm 25 \mu
m}$\,/\,f$_{\rm 60 \mu m}$ $<$ 0.2.
The grouping of LINERs in the 'cool'-starburst-like \ul\ class is consistent with the analysis of \mir\ versus optical
spectroscopic classification performed by
\inlinecite{1999ApJ...517L..13Lutz} discussed in the previous section. 
This, together with the \nir\ colours 
(see Figure 6 in
\inlinecite{2001A&A...379..823Klaas}) 
of
\uls, suggest that the \fir\ emission from LINER-\uls\ is of starburst
origin.
\inlinecite{2001A&A...379..823Klaas} interpreted the emission of 'warm'-\uls\ (type A) to arise from dust
heated to high temperatures directly from the central AGN, which in turn dominates the \mir\ emission. The second SED feature is ascribed
due to thermal re-radiation by dust heated by starburst activity. The
indistinguishable \fir\ SEDs of AGN- and starburst-dominated \uls\
suggests that the
\fir\ emission largely comes from less active or shielded regions that are generally not heated by the AGN \cite{2001A&A...379..823Klaas}. 
Thus \inlinecite{2001A&A...379..823Klaas} proposed a three-stage dust
model consisting of hot AGN heated dust, warm starburst heated dust
(50K$>$T$>$30K) and cold, pre-starburst or cirrus like dust
(30K$>$T$>$10K). The general consensus from \iso\ spectroscopic results of warm \uls\ being AGN-dominated and cold \uls\ starburst-dominated is corroborated by the findings of \inlinecite{2001A&A...379..823Klaas}.
\subsection{\hls\ and the increasing role of AGN at high luminosity}
\label{hls}
The increasing importance of AGN with increasing luminosity \cite{1996ApJ...470..222Shier,1997ApJ...484...92Veilleux} has been
confirmed by \iso\ observations of \uls\
\cite{1998ApJ...505L.103Lutz,2001ApJ...552..527Tran,2002A&A...391..429Charmandaris}. \inlinecite{2001ApJ...552..527Tran}  quantified that the change from
starburst- to AGN-dominated systems occurs at
log$_{10}$(L$_{IR}$)$ \sim $12.4-12.5$L_{\odot}$. While \iso\ results show that
$10^{12}L_{\odot}$ can be powered by star formation alone
\cite{1998ApJ...498..579Genzel,1998ApJ...505L.103Lutz,1999AJ....118.2625Rigopoulou,2001ApJ...552..527Tran},
beyond $10^{13} L_{\odot}$ an AGN contribution seems inevitable. Moreover,
the production of such a luminosity would require star formation rates
$\gsim$ few$\times 10^3$M$_{\odot}$yr$^{-1}$.
Such rates require huge
concentrations of molecular gas to be present (most likely the result
of mergers) and highly ineffective negative feedback
\cite{2003MNRAS.340..813Takagi}.

The prevalence of AGN in \hl\ samples partially arises from a
selection effect. These rare luminous sources have been discovered
with heterogeneous selection methods, most commonly through a
correlation of known quasar or radio catalogues with \iras\ or \iso\
surveys resulting in the known population being biased towards
AGN. Constraining the properties of this sample and full consideration
of the biases involved is unfortunately not possible due to the
paucity of known
\hls. Nevertheless \iso\ observed several hyperluminous galaxies
in a range of programs.

Only a few \hls\ were sufficiently bright to permit
low resolution spectroscopic observations.
Analyses of IRAS 09104+4109
\cite{2001ApJ...552..527Tran} and IRAS F15307+3252
\cite{1998A&A...334L..73Aussel} showed that 
the \mir\ emission predominantly originates from very compact regions
$<$100pc \cite{1997A&A...328L...9Taniguchi} with
lower limits on the size of the emission region of 6pc for F15307
\cite{1998A&A...334L..73Aussel} and $5/\sqrt{m}$ ($m$=magnification) for the
Cloverleaf. These quasars show hot gas at temperatures in
excess of 400K. They exhibit low UIB-to-continuum ratios expected for AGN. For F15307 SED modelling showed that a starburst
component is required to fit this source fuelling 45\% of its
luminosity, although this poses a problem for the
non-detection of CO \cite{2002MNRAS.335..574Verma}. In a sample of
four \hls\ including HyLIGs without clear AGN features, a SED analysis
indicated an AGN was required to explain the IRAS-ISO SEDs
\cite{2002MNRAS.335..574Verma}. One of the highest redshift sources
observed by \iso\ was the powerful radio galaxy 8C 1435+635 (z=4.25). It was
shown to be hyperluminous and massive but the
\ir-mm emission was solely ascribed to star formation and extremely high
star formation rates were implied \cite{1998ApJ...494..211Ivison}.

Several \hls\ are also well known PG quasars that do
not conform to being dust free naked quasars
\cite{2000A&A...354..453Haas}. \inlinecite{2003A&A...402...87Haas}
show a schematic view of the evolution of SEDs encompassing various
types. A significant AGN contribution is seen but PG quasars lack the cooler
extended dust component seen in lower luminosity sources. Patchy or clumpy disk torus models may explain all of the \fir\ emission, without invoking starburst contributions.
\subsection{Starburst components in AGN-hosting galaxies}
\label{stagn}

Is concurrent star formation a requisite to explain the SEDs of QSOs
displaying a \fir\ excess? 
Even the cold \fir-sub-mm emission of AGN dominated sources may be fuelled by
the central obscured active nuclei itself. While \iso\ data cannot
spatially resolve such emission, invoking strong star formation is not
always necessary to describe the SEDs of AGN and ultraluminous quasars (see below).
A clumpy distribution of obscuring material around the black hole
allows high energy photons to escape and heat extended dust (\inlinecite{2002ApJ...570L...9Nenkova},
\inlinecite{2003astro.ph..9040Elitzur}). Alternatively warped, thick and extended disks (on parsec scales)
have been proposed to explain the emission (e.g. \inlinecite{1994MNRAS.268..235Granato}). 

There is clear evidence that some AGN do not host
bolometrically important starbursts. 
For example, the archetypal Seyfert 2
galaxy NGC5252, shows no signs for starburst activity and the bulk of
its \fir\ SED can be explained by the heating of the cold dust by the
general ISRF while a small fraction is ascribed to a very large disk
(kpc scale) surrounding the active nucleus
\cite{2003ApJ...583..689Prieto}.
The \ir\ emission of narrow-line Seyfert 1
galaxies measured by \pht\ can be accounted for by re-radiation of the soft X-ray
component
\cite{1999usis.conf..953Polletta,1999A&A...350..765Polletta}. Similarly, at higher AGN power,
starburst activity is not a necessary requisite to explain the full
SEDs of a large sample of 3CR sources
\cite{2004A&A...421..129Siebenmorgen} or PG quasars \cite{2000A&A...354..453Haas,2003A&A...402...87Haas} for which AGN heating plus extended cold dust suffices (also see \opencite{2005A&A...433...73Rocca}).  In a sample of 22 quasars,
starburst emission does contribute to the \fir\ emission but to a level 
no more than 27\% \cite{2000A&A...362...75Polletta}.  

Thus while concurrent starbursts may be present, \iso\
SEDs of optically classified AGN and quasars indicate that 
they are typically not bolometrically dominant.
The detection of spatially resolved starburst tracers in
the host galaxies of
quasars or AGN-like \uls\ and \hls\ would clarify this
issue.

\subsection{AGN components in starburst-dominated galaxies}

Infrared diagnostics were largely successful in assessing the nature
of most infrared sources.  However, how strong a limit can one place upon the absence of an AGN in the \mir?
At short \mir\ wavelengths $<10$\mn\ the strong VSG continuum related to the star-formation in starburst dominated sources will overshadow the weaker AGN continuum, particularly in Sey 2s where the continuum due to the AGN will suffer high obscuration \cite{2000A&A...359..887Laurent,2000ibp..conf..245Schulz}. This results in an AGN being identified through the \inlinecite{2000A&A...359..887Laurent} diagnostic only if it dominates over star formation.
Galaxies such as IRAS
00183-7111 and Arp 220 are highly obscured even in the \mir\
\cite{1996A&A...315L.133Sturm,2001ApJ...552..527Tran,2004A&A...414..873Spoon},
Arp 220 may be optically thick even in the \fir\
\cite{1997AAS...191.8903Fischer,2001A&A...367L...9Haas,2004ApJ...613..247Gonzalez}). This substantial
obscuration by dust may cause any AGN to be completely hidden \cite{2000ibp..conf..245Schulz}, a fact which is compounded by the possibility that AGN tracers
could in some cases be completely overshadowed by circumnuclear
activity
\cite{1999usis.conf..913Laurent,2003A&A...406..527Dennefeld}. Without spatial resolution, the possibility of detecting extremely obscured or weak AGN in, at least the \mir, appears to be low \cite{2000A&A...359..887Laurent,2000ibp..conf..245Schulz}.
\subsection{Newly detected \uls}

As well as the \uls\ and \hls\ described here which were known prior to
the launch of \iso, several were discovered primarily in \iso\ surveys. Among the many \uls\ detected in the
deep \iso\ surveys, a hyperluminous QSO
\cite{2001MNRAS.327.1187Morel} was discovered in ELAIS, and a highly extincted (A$_V
\sim 1000$) gas and dust rich elliptical was detected in the
\pht\ Serendipitous Survey \cite{2003A&A...402L...1Krause}. One might
expect the population of heavily obscured extremely red objects (EROs)
to be luminous in the \ir. Indeed, ultraluminous emission (and high star formation rates) was
determined from the \iso\ SEDs of the ERO HR10 
with a SED similar to that of Arp220
\cite{2002A&A...381L...1Elbaz} and a sub-mJy radio ERO from the Phoenix survey (where coeval
star formation and AGN are probable
\cite{2001ApJ...559L.101Afonso}) through targeted observations. 
Tentative \ir\ detections of a gamma ray burst were also reported to be at \ul\ luminosities \cite{1999A&AS..138..459Hanlon,2000A&A...359..941Hanlon} originating from either the host galaxy or the GRB \ir\ afterglow.
\subsection{Evolution of \uls}
\label{evolul}
The basic similarity in luminosity and space density of \uls\ and
quasars, and the fact that unbiased samples of \uls\ contained both
star forming and AGN members, led to the suggestions that the two
populations are linked through evolution (\inlinecite{1988ApJ...328L..35Sanders}, also see \inlinecite{1996ARA&A..34..749Sanders} and references therein). The example of the composite
galaxy \n985 is used to discuss possible \ul\ formation and
destruction scenarios
\cite{2002ApJ...566..682Appleton}.
Some popular evolutionary
models include the following that were recently well summarised by
\inlinecite{2003MNRAS.340..289Lipari}:\\ 
(1) merger - giant shocks -
super-starbursts + galactic winds - elliptical galaxies \cite{1999Ap&SS.266..321Joseph} \\ (2) merger - H$_2$ inflow (starbursts) - cold ULIRGs - warm
ULIRGs+QSOs - optical QSOs \cite{1999Ap&SS.266..331Sanders}
\\ (3) merger/s -
extreme starburst + galactic wind (inflow + outflow) - FeII/BAL
composite IR-QSOs - standard QSOs and ellipticals \cite{2003MNRAS.340..289Lipari} \\

\subsubsection{Merger Stage and Luminosity}
Merging activity that initiates associated starbursts leading to the
ultraluminous behaviour in the early stages is common to all of the
scenarios above.
Thus the merger stage should be related to the
luminosity evolution of \uls. Furthermore, if quasars are the end
product of \ul\ evolution and the merger stage indicates an evolutionary
status, then more advanced mergers should appear more AGN-like. 

The near- to mid-infrared energy distributions of \uls\ are dominated by compact
nuclear regions \cite{2002A&A...391..429Charmandaris}. This confirms
results from high resolution ground-based observations
\cite{2000AJ....119..509Soifer} that show 30-100\% of the \mir\ flux
can be accounted for by the central 100-300pc.
Based on \iso\ imaging of a merging sequence of active galaxies,
including starbursts, \ls\ and
\uls, \inlinecite{2001ApSSS.277...55Charmandaris} suggest an evolutionary sequence that is traced by the
15\mn/7\mn\ colour that monotonically increases with star formation
activity, from 1 (quiescent) to 5 (extreme), as galaxies evolve from
pre-starburst to the merging starburst phase. 
The ratio decreases
again to 1 for the post starburst phase in evolved merger remnants
[cf local less luminous post-starburst colours of \n7714 \cite{2000A&A...360..871Ohalloran} and \n1741 \cite{2002ApJ...575..747Ohalloran}]  
The fact that the extra-nuclear activity typically contributes only a
minor fraction to the bolometric luminosity of \uls\ (in comparison to
\ls\ \cite{1999ApJ...511L..17Hwang}) implies that simple scaling in mass and luminosity does not link
the two populations, but evolutionary differences (such as merger
stage, or presence of an AGN) could play a role. It is unclear whether this hardening of the \mir\ continuum is
due to an AGN contribution.
The amount of molecular gas may also affect the spectral feature
differences seen between the populations, such as the presence of ice
features
\cite{2002A&A...385.1022Spoon}. Increasing compactness of sources with increasing merger
stage shown by \inlinecite{1999ApJ...511L..17Hwang} implies that
compact \uls\ are in a more advanced merger state than \ls.

However, based upon a comparison of the NIR morphologies of merging
\uls\ and \ir\ spectroscopic tracers,
\inlinecite{1998ApJ...505L.103Lutz} and \inlinecite{1999AJ....118.2625Rigopoulou} do not find
any relation of increasing AGN-like activity with merger stage. In
addition \uls\ appear to not show a trend of increasing luminosity
with merger stage again contrary to expectations of these evolutionary
scenarios
\cite{1999AJ....118.2625Rigopoulou}. Indeed
\inlinecite{1999AJ....118.2625Rigopoulou} suggest that early merger stage
\uls\ may be exhibiting maximal luminosities (also see \opencite{2001ApJ...559..201Murphy}). 
Moreover, the mass of the warm, cold and very
cold dust components found in \ul\ SEDs show no relation to
merger stage  \cite{2001A&A...379..823Klaas}. The evolution or merger-stage may not be traced well simply by nuclear separation alone as a merger stage indicator. More complex differences
in the mass distribution and dynamics of
\uls\ more likely influence the luminosity and appearance of AGN.

\subsubsection{Dust evolution in quasars}
If \uls\ are related to quasars as the dust obscured members of that class,
understanding the dust properties of quasars also serve to test
the evolutionary connections between quasars and \uls. 
In the
\inlinecite{1988ApJ...328L..35Sanders} scenario a quasar is preceded
by a dusty \ul-phase and as a result it is unlikely that the obscuring dust 
will suddenly disappear, some traces in the \ir\ should remain.
\inlinecite{2003A&A...402...87Haas}
investigated a sample of 64 PG quasars and identified a range of SED shapes that could mark a transition between states. 
The SED analysis showed that the central regions of PG quasars suffer low
extinction (A$_{V}$\,$<$\,0.3) with an optical slope
$\alpha$$_{opt}$ that is independent of \ir\ properties like the
near- to mid-\ir\ slope $\alpha$$_{IR}$. Therefore, with regard to
the unification schemes a nearly face-on view onto the PG
quasars was assumed.
The observed variety of SEDs were grouped into
physically meaningful classes, that reflect the amount and
distribution of the reprocessing dust around the AGN according to
evolution of the
quasar
\cite{2003A&A...402...87Haas}.
Their new evolutionary
scenario begins with dissipative cloud-cloud collisions and angular momentum
constraints that create the organisation of dust clouds into a
torus/disk like configuration. Starburst activity is initiated giving rise to a
\fir\ bump but it becomes increasingly overpowered by the
central AGN which results in increasing \mir\ emission and steepening of the mid- to \fir\
spectral slope. As the AGN continues to grow, the \mir\ remains high and
the mid- to \fir\ slope decreases until the black hole begins to starve and a a decline
in the \mir\ and \fir\ emission is observed. Figure 12 in \inlinecite{2003A&A...402...87Haas} depicts this scheme,  the SEDs are
arranged along the expectations for such an evolutionary scheme. 

\subsubsection{Relation of broad absorption line quasars to \uls} 
Broad
absorption line quasars are a rare class of the optically selected quasar
population ($\sim$12\%) but have a higher frequency in \fir\ luminous, strong
FeII and weak [OIII] emission line selected samples. 
Both orientation and age arguments are consistent with the origin of
the broad absorption features with no clear resolution which is
the most likely cause. 
\inlinecite{2000NewAR..44..559Elvis} suggest the broad absorption arises from viewing a single funnel-shaped outflow of absorbing material directly down the flow.
It has also been proposed that these sources are quasars
viewed at an orientation
directly along the radial surface of the torus, where winds along the surface give rise to the observed absorption.
Alternatively they may
be a young phase in quasar evolution. The latter together with the
high occurrence of BALQSO in \iras\ selected samples suggests that they
may be a transition stage between \uls\ and QSOs
\cite{1993ApJ...413...95Voit,1994ApJ...436..102Lipari,1996AJ....112...73Egami,2001ApJ...555..719Canalizo,2003MNRAS.340..289Lipari}. In this scenario very young QSOs are ejecting their gaseous envelopes at
very high velocity close to the initiation of the active phase of the black hole \cite{1984ApJ...282...33Hazard}.

Two BALQSOs were serendipitously
discovered in \iso\ surveys: the FeLoBAL\footnote{FeLoBALs show absorption lines from
many excited states of Fe II and having low ionisation lines and are
heavily absorbed in X-rays}
ISO J005645.1-273816 was discovered in an
ISOCAM distant cluster survey \cite{2002A&A...389L..47Duc}. 00 37 14.3
-42 34 55.6 was discovered in ELAIS
\cite{2001ApJ...554...18Alexander}. For the former, high absorption and hot dust emission are implied by the high \mir\ to UV ratio which may indicate the \mir\ is an efficient means to detect LoBALQSOs \cite{2002A&A...389L..47Duc}.
 The sources 00 37 14.3 -42 34 55.6 was the only BALQSO then known to have both \ir\
and hard X-ray detections. While it has a similar hard X-ray/MIR flux ratio
to high redshift quasars, it differs from low-redshift BALQSOs.
\citeauthor{2001ApJ...554...18Alexander} attribute this to absorption. As a consequence they suggest that
absorbed BALQSOs at high redshift should be easily detected in
the hard X-ray bands. In addition the only radio-loud BALQSO known,
1556+3517, was observed to investigate the relation between radio-loud
quasars and the BALQSO phenomenon. Mid-\ir\ detections implied
4M$_{\odot}$ of dust lie along the line of sight which most likely
arises from the molecular torus, the NLR, or a recent starburst rather
than from the BALQSO wind itself
\cite{1998A&A...331..853Clavel}.
\iso\ data confirms that even in this diverse, small sample, 
the BALQSO phenomenon and obscuration (i.e. \ir\ emission) are
connected. \inlinecite{2003MNRAS.340..289Lipari} speculate on 
an
evolutionary connection between BALQSOs and \uls.

\subsection{Summary}
Evaluating which of the three scenarios mentioned earlier in this
Section is most realistic will continue to be addressed by future
missions. However, the \iso\ results clearly 
suggest that a simple \inlinecite{1988ApJ...328L..35Sanders} evolutionary scenario of a merger triggered
starburst event with an evolving obscured- naked quasar scenario to be
unlikely. Rather, it seems that the nature of the diverse \ir\ luminous
\uls\ and \hls\ has a far more complex history. Whether a starburst or
an AGN dominates the bolometric luminosity of a \ul\ is most probably
determined by local conditions, evolution and obscuration by
dust. Nearby \uls\ are mostly starburst dominated systems with 'cold' colours. This cold
fraction decreases at higher luminosities where AGN
contributions become prevalent and galaxy SEDs are 'warmer'. \uls\ are
often linked to high-redshift (z$\sim$2) (sub-)mm emitting galaxies which comprise
a significant fraction of the \ir\ background and contribute significantly to the star formation density at that epoch. 
%Conversely, local luminous \ir\ galaxies contribute less than 1\%. 
If \uls\ and sub-mm galaxies are members of the same population 
%then strong evolution is implied.
%If this is proven true, 
then strong evolution is implied since
the locally unimportant \uls, that constitue less than 1\% of the local star-formation density, must dominate the \ir\ number
counts at higher redshifts \cite{2003Sci...300..270Elbaz}. 
The strongly evolving z$\lsim$1.5 \ls\ discovered in \cam\
surveys (see Elbaz et al. and Oliver et al. in this volume) that
resolve 10-60\% of the far-IR-sub-mm background may provide a step
between.

\section{Active Galaxies as seen by ISO: a summary}
The Infrared Space Observatory has produced a wealth of information
regarding the infrared emission properties of the full range of active
galaxies. Greatly enhancing the IRAS legacy, it has provided a
physical basis upon which the next generation of infrared telescopes
such as Spitzer, SOFIA, ASTRO-F and Herschel will build. Most of \iso's
power was due to the improved sensitivity and extended wavelength
coverage. The addition of high resolution spectroscopy and development
of new diagnostic diagrams have enabled us to probe the ongoing physical conditions of the ISM in active galaxies and to quantitatively assess
the presence and bolometric contribution of star formation or an AGN
in infrared luminous sources. Analysis of fine structure lines provided
an excellent set of diagnostic tools that will be better populated and
applied to a wider variety of sources in future missions with even
higher sensitivity. The weakness of the sulphur fine structure lines in starburst galaxies 
and of the cooling line [CII]158\mn\ in the most luminous active galaxies suggests that other strong lines
(e.g. [NeII]12.8\mn) should be the are preferred indicators of star
forming objects. The finding of \iras\ that galaxies with "cold"
\ir\ colours are starburst dominated and "warm" are AGN dominated
was broadly confirmed with the new \iso\ data. Although, it was shown that
spatially resolved spectroscopic information is necessary to address
the issue of the origin of the \fir\ emission in AGN and the role
of concurrent star formation. \iso\ results on AGN in general
support unification scenarios, but suggest that a clumpy distribution of
obscuring material around the active nucleus may be a better
representation than the commonly assumed torus. While the majority of
\uls\ are starburst dominated, it now appears that an increasing 
fraction of AGN at higher infrared luminosities is present. Mergers are 
indeed the trigger of \ul\ activity, but the simple generic evolution of 
a luminous system from a \ul\ to an optical-QSO phase appears unlikely with more
local factors influencing the presence of an AGN. With
increases in sensitivity and spatial resolution, it is certain that
the suite of current and forthcoming infrared telescopes will enhance
our knowledge of the obscured active universe by probing fainter and
to higher redshifts as well as in more spatial detail.

\begin{acknowledgements}
We would like to thank Matthias Tecza for his comments on this text. We are also grateful to Eckhard Sturm for providing the SWS+LWS spectra shown in Figure 1 and for discussion. This work was supported by the DLR grant FKZ: 50QI 0202.
\end{acknowledgements}

\bibliographystyle{klunamed}
%
%\bibliography{/home/verma/MYPAPERS/all}

\label{lastpage}
\end{article}

\end{document}